\documentclass[]{aastex63}
\usepackage[utf8]{inputenc}
\usepackage{amsmath}
\usepackage{breqn}
\usepackage{listings}
\usepackage{xcolor}
\usepackage{hyperref}
\hypersetup{linkcolor=red,citecolor=blue,filecolor=cyan,urlcolor=blue}

\lstdefinestyle{mystyle}{
    commentstyle=\color{codegreen},
    keywordstyle=\color{magenta},
    numberstyle=\tiny\color{codegray},
    stringstyle=\color{codepurple},
    basicstyle=\ttfamily,%\footnotesize,
    breakatwhitespace=false,         
    breaklines=true,                 
    captionpos=b,                    
    keepspaces=true,                 
    numbers=none,                    
    showspaces=false,                
    showstringspaces=false,
    showtabs=false,                  
    tabsize=2
}
\usepackage[caption=false]{subfig}

\graphicspath{ {./Figures/} }

\newcommand{\dd}{\mathrm{d}}
\newcommand{\cat}{\ensuremath{\mathcal{C}}}
\newcommand{\unit}[1]{\mathrm{#1}}
\newcommand{\pc}{\unit{pc}}
\newcommand{\mas}{\unit{mas}}
\newcommand{\Mag}{\unit{mag}}
\newcommand{\vq}{\ensuremath{\mathbf{q}}}
\newcommand{\BmR}{\ensuremath{(B-R)}}
\newcommand{\BmG}{\ensuremath{(B-G)}}
\newcommand{\GmR}{\ensuremath{(G-R)}}
\newcommand{\mG}{\ensuremath{G}}
\newcommand{\aMG}{\ensuremath{M_G}}
\newcommand{\modelpars}{\ensuremath{\boldsymbol{\varTheta}_\mathrm{mod}}}
\newcommand{\plxSN}{\ensuremath{\frac{\varpi}{\sigma_\varpi}}}
\newcommand{\eplxSN}{\ensuremath{ {\overline{\mathrm{S/N}_\varpi} } }}
\newcommand{\plxSNlim}{\ensuremath{ {\overline{\mathrm{S/N}}_{\varpi ,\mathrm{min}}} } }

\newcommand{\lcf}{\ensuremath{\Phi_0(M,c)}}
\newcommand{\ndens}{\ensuremath{\hat{n}(\mathbf{x})}}
\newcommand{\ndenspars}{\ensuremath{\hat{n}(\mathbf{x}\mid\modelpars)}}
\newcommand{\ndenssky}{\ensuremath{\hat{n}(\mathbf{x}_{\rm sky},d)}}
\newcommand{\paf}{\ensuremath{p_\mathrm{af}}}
\newcommand{\vmax}{\ensuremath{V_\mathrm{max}}}
\newcommand{\veff}{\ensuremath{V_\mathrm{eff}}}

\renewcommand{\twocolumngrid}{\smallskip}
\sloppypar
\shorttitle{selection functions in data modeling}
\shortauthors{rix et al.}

\newcommand{\cca}{Center for Computational Astrophysics, Flatiron Institute, 162 Fifth Ave, New York, NY 10010, USA}
\newcommand{\ccpp}{Center for Cosmology and Particle Physics, Department of Physics, New~York~University, 726~Broadway, New~York, NY 10003, USA}
\newcommand{\mpia}{Max-Planck-Institut f\"ur Astronomie, K\"onigstuhl 17, D-69117 Heidelberg, Germany}
\newcommand{\leiden}{Leiden Observatory, Leiden University, Niels Bohrweg 2, 2333 CA Leiden, Netherlands}
\newcommand{\cambridge}{Institute of Astronomy, University of Cambridge, Madingley Road, Cambridge CB3 0HA, United Kingdom}
\newcommand{\oxford}{Magdalen College, Oxford University, Oxford OX1 4AU, United Kingdom }
\newcommand{\torino}{INAF - Osservatorio Astrofisico di Torino, Strada Osservatorio 20, Pino Torinese 10025 Torino Italy}
\newcommand{\monash}{School of Physics and Astronomy, Monash University, VIC 3800, Australia}

\begin{document}
\title{Selection Functions in Astronomical Data Modeling, \\
with the Space Density of White Dwarfs as Worked Example.}

\author[0000-0001-5996-8700]{Hans-Walter Rix}
\affil{\mpia}

\author[0000-0003-2866-9403]{David W. Hogg}
\affil{\mpia}
\affil{\cca}
\affil{\ccpp}

\author[0000-0002-7521-6231]{Douglas Boubert}
\affil{\oxford}

\author[0000-0002-7419-9679]{Anthony G.A. Brown}
\affil{\leiden}

\author[0000-0003-0174-0564]{Andrew Casey}
\affil{\monash}

\author[0000-0002-1777-5502]{Ronald Drimmel}
\affil{\torino}

\author[0000-0001-5620-2232]{Andrew Everall}
\affil{\cambridge}

\author[0000-0001-9256-5516]{Morgan Fouesneau}
\affil{\mpia}

\author[0000-0003-0872-7098]{Adrian~M.~Price-Whelan}
\affil{\cca}

\begin{abstract}\noindent
 Statistical studies of astronomical data sets, in particular of cataloged properties for discrete objects, are central to astrophysics.  One cannot model those objects' population properties or incidences without a quantitative understanding of the conditions under which these objects ended up in a catalog or sample, the sample's \emph{selection function}. As systematic and didactic introductions to this topic are scarce in the astrophysical literature, we aim to provide one, addressing generically the following questions: What is a selection function? What arguments {\vq} should a selection function depend on? Over what domain must a selection function be defined? What approximations and simplifications can be made? And, how is a selection function used in `modelling'? We argue that volume-complete samples, with the volume drastically curtailed by the faintest objects, reflect a highly sub-optimal selection function that needlessly reduces the number of bright and usually rare objects in the sample.
 We illustrate these points by a \href{https://github.com/gaia-unlimited/WD-selection-function}{worked example}, deriving the space density of white dwarfs (WD) in the Galactic neighbourhood as a function of their luminosity and Gaia color, $\Phi_0(\aMG,\BmR)$ in $[\Mag^{-2}~\pc^{-3}]$. We construct a sample $\cat$ of $10^5$ presumed WDs through straightforward selection cuts on the Gaia EDR3 catalog, in magnitude, color, parallax, and astrometric fidelity,
 $\vq=(\mG,\BmR,\varpi,\paf)$. We then combine a simple model for $\Phi_0$ with the effective survey volume derived from this selection function $S_\cat(\vq)$ to derive a detailed estimate of $\Phi_0(\aMG,\BmR)$ is robust against the detailed choices for $S_\cat(\vq)$. This resulting white dwarf luminosity-color function $\Phi_0(\aMG,\BmR)$ differs dramatically from the initial number density distribution in the luminosity-color plane: by orders of magnitude in density and by four magnitudes in density peak location.
  
\end{abstract}

\section{Selection Functions in Astronomy}
Statistical studies of astronomical data sets or catalogs are central to many, if not most, aspects of astrophysics. They usually entail \emph{making a model} of some of the cataloged quantities that characterize (usually discrete) sets of objects, and constraining that model by asking quantitatively whether the data in the catalog match model expectations. This requires that one understands under which circumstances an object would have had a chance to be in the catalog, or in a sub-sample drawn from a catalog.
This understanding can be captured in a \emph{selection function} (or \emph{selection probability}). The probability of an object to be in the catalog depends on, for example, the detection efficiency of the observational survey from which the catalog was derived as well as on choices made during the construction of the catalog, such as removing potential entries deemed to be of insufficient `quality'. 

Implicitly, the issue of selection functions in astronomy has been around for as long as there have been astronomical catalogues, with perhaps the first articulations given in classic textbooks such as \cite{TrumplerWeaver1953}. The selection function of a sample is closely related to the effective or maximal survey volume \vmax\ of a catalog (a spatial or a generalized parameter-space volume), a concept introduced quantitatively by \cite{Schmidt1968} in the context of seminal work on quasars. The concept of a selection function has also been linked to the concepts of \emph{selection effect} or \emph{selection bias}; we deem those concepts to be more nebulous, as they have been used both as a synonym for the selection function itself, and for biased results arising from ignoring important aspects of the selection function in an analysis.

The concept of a selection function is very widely used in contemporary astrophysics: ADS lists $700$ instances of it in the refereed publications of the year 2020 alone. Yet, didactic expositions of selection functions -- definitions, worked examples, best-practices guidance --  are hard to find in the astrophysical literature (see \citealp{BR13,Wojno2017,Stellar_Inventory_Bovy_2017,BE20,Smart20} for notable recent exceptions).  This paper aims to fill this gap by providing an exposition of the conceptual and practical issues that arise when devising and applying a selection function. 

We will use the Gaia catalog \citep{Gaia,eDR3} -- one of the most extensive, multi-dimensional all-sky catalogs of discrete astronomical objects -- as a backdrop and input to our worked example. However, we stress that the basic formulation and many aspects of the suggested best-practices should have far broader applicability. Framing these issues in the Gaia context is based on two considerations. First, with precision measurements of $10^{6}{-}10^{9}$ objects, statistical analyses of Gaia data will rarely be limited by the sample size (and its Poisson variance), and often not by the individual measurement precision. % ARC: precisions? I dropped the 's'. Or maybe I mis-understood something.
Instead, modelling will be limited by the precision and the incorporation rigor of the selection function. Second, awareness of the central role of a selection function and of established techniques to implement it in modelling are perhaps not as widespread across all aspects of `Gaia science' as they are in cosmological large scale structure \citep[e.g.][]{Cole2005}, or in gravitational wave detection.

The remainder of the paper is structured as follows: we aim to summarize selection function basics in Section~\ref{Basics}; we illustrate these with a \emph{worked example} in Section~\ref{M_c_WD}, deriving the white dwarf (WD) \emph{luminosity--color function} (LCF) from Gaia data, the solar neighbourhood space density of WDs, in $\pc^{-3}$, as a function of their absolute magnitude \emph{and} color. A python notebook for this worked example can be found \href{https://github.com/gaia-unlimited/WD-selection-function}{here}. We then lay out a number of further issues that should be considered when applying and deriving selection functions in Section~\ref{sec:discussion}; for most, their detailed resolution is beyond the scope of this paper.

Some readers may prefer to start by seeing a concrete example of how to model data including a selection function, before considering the broad guidance in Section~\ref{Basics}.
We encourage those readers to skip forward to Section~\ref{sec:SF_in_data_model_comparison} and then go to Section~\ref{M_c_WD}, before returning to the rest of Section~\ref{Basics}.

\section{Selection Function `Basics'}\label{Basics}

In this Section we give a general introduction to the concept and use of selection functions in astrophysics, addressing:  What is a selection function, what are desirable properties for it, what is its role in modeling?

It is a common situation in astronomy that we have a model for the physical properties of discrete objects, say stars, and we want to test this model (or find its best-fit parameters) through a comparison with data. And it is quite likely that one of astronomy's vast catalogs lists observational constraints (fluxes, color, etc.) on such objects; one then selects a pertinent subset of such objects and fits a model to them. 

\emph{What is a selection function?\  }
There are several ways to look at it:  One may view the selection function, $S_\cat(\vq )$, as the {\it probability} that an object with attributes {\vq} will be contained within a catalog or sample  {\cat} under consideration; we use the subscript {\cat} for the selection function as a constant reminder that it is `for a given catalog or sample {\cat}'. 
Operationally, a catalog {\cat} in the current context is simply a list that specifies attributes {\vq} for a set of discrete objects. The full set of these catalog attributes can and often will be more extensive than the set of {\vq} that enter the selection function or are being modeled. 

The most common use of selection functions is in `modelling' data sets, based on some model
family, $\mathcal{M}(\vq\mid\modelpars)$ parameterized by $\modelpars$. There
the selection function may also be viewed as the multiplicative link between the probability density predictions of  $\mathcal{M}(\vq\mid\modelpars)$ for the quantities {\vq}, and the {\it expected} catalog-incidence (if the model were correct), $\dd\Lambda_\cat(\vq)$, of these quantities:
\begin{equation}
   \dd\Lambda_\cat(\vq) =  \mathcal{M}(\vq\mid\modelpars)\  S_\cat(\vq )~\dd\vq.
   \label{eq:SFdefinition}
\end{equation}
% ARC: It may be verboten, but I have added a \, so that the arrow above q is distinct from the quotation.
Note that while the model prediction is a probability density with units `per $\dd\vq\,$', the selection function is simply a unitless probability (function), bounded between zero and unity\footnote{The upper bound of 1 may in rare cases be exceeded, if a catalog construction leads to a finite probability that one object leads to multiple (unlinked) catalog entries.}. We will lay out below the question of which arguments {\vq} of the selection function are suitable and necessary.

\emph{When does one need a selection function?\ }
Broadly, we need to determine and apply a selection function whenever we want to answer a question or constrain a model through data comparison, and when that model predicts densities, rates, or other incidences for objects with certain characteristics (that are reflected in `observables'). Note that selection functions are not only needed when analyzing large sets of catalog entries, but just as much when an extensive search for elusive objects has yielded one specimen (or even none); after all, the predicted catalog incidences for a physical model in Eq.\ref{eq:SFdefinition} can well be $\Lambda_\cat(\vq)< 1$. 

\subsection{How does one construct a selection function?}
It is easy to state that all one needs for stringent modelling of objects in {\cat} is a sensible model family $M(\vq\mid\modelpars)$ and a selection function $S_\cat(\vq)$ in the sense of Eq. \ref{eq:SFdefinition}; but this does not address how to devise a good selection function and its resulting sample {\cat}. 

In the context of large, contemporary astronomical data catalogs (Gaia, PanSTARRS, 2MASS, WISE, GALEX, eROSITA, ...) it is rare that anyone aims to build a model that tries to constrain the physical properties of all objects in the entire \emph{parent} catalog at once: usually such catalogs encompass objects of very different physical natures, from say AGN to White Dwarfs, for which it makes little sense to build a model simultaneously. Instead it is most common to model only subsets of objects from the parent catalog, constructed foremost by cuts or selections in properties {\vq} or in aspects of `data quality'.

\begin{figure}[ht]
    \centering
    \includegraphics[width=0.6\textwidth]{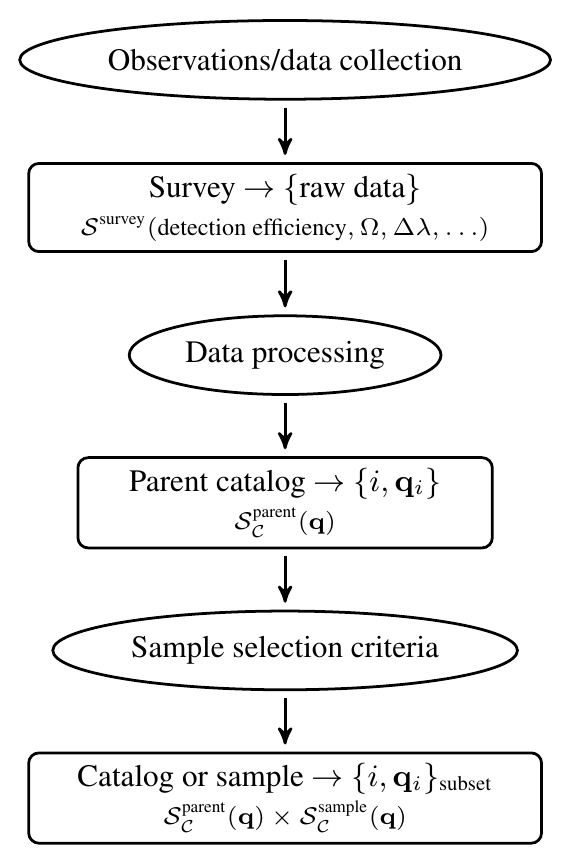}
    \caption{Schematic of all the factors that sequentially set the selection function $S_\cat(\vq)$ of a sample of discrete astrophysical objects to be modelled. $S_\cat(\vq)$ describes the probability that such an object with observable properties $\vq$ will enter a sample. Viewed end-to-end, this starts with the overall experiment (say, the Gaia mission) and its \emph{detection efficiency} of astrophysical sources, $S^{\rm survey}$. After data processing, this results in a parent catalog (say, the Gaia EDR3 catalog), whose \emph{completeness}, $S_\cat^{\rm parent}(\vq)$ must be characterized. Astrophysical models for some class of objects can then be constrained by comparison with catalog data. But in practice only a (often tiny) subset of objects in the entire parent catalog 
    will be modelled. Commonly, these are objects that represent some particular class of astrophysical objects, say white dwarfs, or QSO, etc. Such sub-samples are typically defined
    through a set of selection `cuts', $S_\cat^{\rm sample}(\vq)$. 
    In the end, the individual selection function factors are multiplied and summarised in the overall \emph{selection function},
    $S_\cat(\vq)$. In practice, the parent catalog completeness, 
    $S_\cat^{\rm parent}(\vq)$ is often the same `given' across many modelling applications, while the sample selection 
    $S_\cat^{\rm sample}(\vq)$ 
    will always be tailored to the astrophysical case at hand.
    }
    \label{fig:selection-catalog-schematic}
\end{figure}

For most applications it therefore makes sense to think of the construction of a selection function as consisting of two parts: the first is to characterize the \emph{detection efficiency} of the underlying experiment and the resulting \emph{completeness} of the parent catalog; the second is the definition of the sub-sample to be modelled through selection on cataloged properties {\vq}. Figure~\ref{fig:selection-catalog-schematic} presents a schematic of this multi-step process towards a `final' or `total' selection function for modelling, which we can use as a guide throughout.
Any step in Figure~\ref{fig:selection-catalog-schematic} towards constructing a selection function reflects inevitably a number of choices: in the experimental design leading to the parent catalog, or scientific choices in selecting a sub-sample from it for the astrophysical problem at hand. Therefore, it may be a formidable task to construct $S_\cat(\vq )$ and understand its fidelity.

Figure~\ref{fig:selection-catalog-schematic} makes explicit that the parent catalog is in practice derived from a survey through some form of processing to turn raw data into the catalog quantities {\vq}. The survey also has a selection function which is determined by the sky coverage, the sensitivity of the telescope-instrument combination, wavelength range, etc. In practice most scientific analyses start from the catalog entries {\vq}, and in the rest of this paper we only discuss the selection function of the parent catalog. We assume that this selection function implicitly accounts for the survey selection function, where for example the sensitivity limit, in combination with a minimum signal to noise ratio needed for the data processing, is translated to a magnitude limit for the parent catalog.

We start by discussing the inherent `completeness' 
$S^{\rm parent}(\vq)$ of the overall parent catalog \citep[e.g.][for Gaia DR2]{BE20}, quantifying the probability that a source with observable characteristics {\vq} is included in the parent catalogue. There are two separate (but non-exclusive) paths to determine $ S^{\rm parent}(\vq) $. The most straightforward path is through knowing some `ground truth', a complete and sufficiently extensive set of objects whose properties we know from external empirical information; to be useful the {\vq} of the ground-truth sample must be such that some will and some won't end up in the parent catalog. Such information can come e.g. from deeper or higher resolution data over restricted survey areas. The parent catalog completeness is then constrained by asking which $S^{\rm parent}(\vq) $ makes the actual catalog entries among the ground-truth sample a likely outcome, as a function of {\vq}.
But very often sufficient ground truth is not known. Then $S^{\rm parent}(\vq) $ must be constructed from an understanding of the overall experiment
and of the data processing that leads to catalog entries. In practice, implementing such an approach rigorously requires considerable effort. For example, hardly any survey is simply flux-limited. In the case of the Gaia mission, $S^{\rm parent}(\vq) $ depends in complex ways on sky position, both because of Gaia's scanning law and because of source crowding. But both aspects are known and can be accounted for, and indeed a parent catalog selection function for Gaia has been derived by \citet{BEH20a,Boubert21} using Binomial statistics to infer the detection efficiency \citep{BE20}. In that application, the resulting $S^{\rm parent}(\vq) $ made no assumption about the ground truth but solely relied on an understanding of the experiment.

While deriving $S^{\rm parent}(\vq)$ may be hard, it can then be used unchanged for basically all science analyses based on this parent catalog. And it is common that subsequent sample cuts $S_\cat^{\rm sample}(\vq)$ keep only {\vq} that stay well away from the survey's detection limit, where the approximation $S^{\rm parent}(\vq) = {\rm const.} \approx 1$ is sensible.

%First, consider the inherent `completeness' of the overall parent catalog \citep[e.g.][for Gaia]{BE20}. This requires an estimate of the `ground truth', a complete set of objects, which can either be estimated from models or from external empirical information (such as deeper data or higher resolution data over restricted areas, etc..). In most cases, a rigorous and detailed assessment of the parent catalog's selection function, $S^{\rm parent}(\vq)$, also requires a good understanding of the experimental details of the data taking and analysis that led to the parent catalog. \com{RD et al}{Expand a bit what "completeness" of the survey or of the parent catalog means. Terminology: survey, or parent catalog? {\bf AE or DB: can you add a few sentences here? }}

We now turn our attention to the second step (see Fig.~\ref{fig:selection-catalog-schematic}), present in almost all astrophysical modelling of catalog data: devising a sub-sample $\cat$ from the parent catalog that encompasses only the objects to be compared to the model, through a suitable choice of $S^{\rm sample}_\cat(\vq)$. Generally, $S^{\rm sample}_\cat(\vq)$, will select $\cat$ only on the basis of a subset  of the parent catalog's attributes, such as sky position, $\mathbf{x}_{\rm sky}$, magnitude, color, parallax ($\varpi$), etc... . But it is also common and sensible to include {\it signal-to-noise} of attribute estimates (such as \plxSN) or `error flags' among those {\vq}, as we will discuss below. 
In general, the final selection function of a selected catalogue $\cat$ from such multi-step procedures can be captured as the result of multiplications:
\begin{equation}
    S_\cat(\vq)=S^{\rm parent}_\cat(\vq)~\cdot ~ S^{\rm sample}_\cat(\vq),
    \label{eq:SF_factorization}
\end{equation}
as per Figure~\ref{fig:selection-catalog-schematic}.
In this context the terminology `selection cut' does not imply that $S^{\rm sample}_\cat(\vq)$ is either 1 or 0. A probabilistic but controlled sub-sampling of the full parent catalog within the domain where $S^{\rm sample}_\cat(\vq)$ has support may well lead to a perfectly well-behaved selection function. If one chooses to model a random fraction $f$ of objects that satisfy $S_\cat(\vq)$ one simply has to use
$S^'_\cat(\vq) \equiv  f\cdot S_\cat (\vq)$

\subsection{What should be the arguments of a selection function?}
Generally, the selection function should be a function of the minimal set of object attributes {\vq} that describes any object's probability to enter {\cat} with sufficient precision and accuracy (which depends on the science question). This broad conjecture deserves some elaboration and qualification:

\noindent$\bullet$\   For almost all modelling -- involving integrals over `parameter space' --  it is important that the selection function describes such probabilities for both actual objects and for arbitrary, or counter-factual, objects that may have postulated attributes {\vq} not represented in the {\cat}. For instance, the selection function of a magnitude-limited survey must return 0 for any counter-factual object fainter than the magnitude-limit, of which there are none in the catalogue. More generally, we must know the selection function beyond the domain in {\vq} where we actually have cataloged objects.
    
    % HOGG SAY: I don't like the _\cat subscripts on the q values. Do we need to subscript q as well as S? It isn't obvious why. Either explain that in the text or drop it.
    
    % HOGG SAY: Let's use \, (thin space) not \times as a multiply operator.
    
\noindent$\bullet$\  Formally, $S_\cat$ may be a function of more arguments, say of the full set of cataloged attributes $\vq_{C}\equiv ( \vq , \vq_{\rm irr})$, where the set of irrelevant attributes $\vq_{\rm irr}$ is (quasi-)defined by $S_\cat^{\rm full}(\vq_{C}) = S_\cat(\vq )\,  S_\cat^{\rm irr}(\vq_{\rm irr})$ with $S_\cat^{\rm irr}(\vq_{\rm irr})\equiv 1\, \forall\, \vq_{\rm irr}$. Note that $S_\cat$ is and remains unitless, irrespective of the number of arguments. In photometric sample selection, magnitudes and colors are manifestly selection-relevant {\vq}, while photometric selection may be `blind' to  radial velocities $v_{rad}$ or proper motions $\boldsymbol{\mu}$ (if not too large), even if they were listed in the catalog; then $v_{rad}$ and $\boldsymbol{\mu}$ would be part of $\vq_{\rm irr}$.

\noindent$\bullet$\  If possible, the {\vq} should represent simple `observables'; that is, the {\vq} should be quantities such as sky positions, magnitudes, colors or parallaxes, rather than model-derived quantities such as \emph{intrinsic colors}, \emph{ages}, \emph{bolometric luminosities},  \emph{temperatures} or \emph{distances}. However, if one looks closely, one realizes that the distinction between observables and model-dependent quantities is not well defined: technically, every catalog entry reflects some form of (data) model for a true observable.

%\com{MF}{After this list, I cannot stop thinking about the part that we do not see: if you have a magnitude limited sample, it's probably not a sharp cut, hence there are assumptions where the selection function is not unity. I cannot find an answer in this list. In other words, S is conditioned on things that we do not measure in our catalog.}   [HWR: adressed above]

    So, the following guidance may be more practicable:\\ \textbf{\textit{The selection function should depend on (a minimal set of) catalog attributes that can be predicted by the model and causally determine the catalog membership probability to sufficient precision.}} \\
    Model predictability matters: for example, young stellar objects can be effectively selected by their optical variability (in conjunction with other criteria). Yet, only a subset of young stellar objects vary and there is no good model that describes which vary and by how much, making quantitative modelling (say, of the age-mass distribution) of variability-selected young stellar objects complex or even impossible. 
    
There is a subset of common catalog attributes that deserve special attention in this context: \emph{data quality flags} and \emph{signal-to-noise cuts}, or cuts on the quoted measurement uncertainties (not the mean estimates). 

% HOGG SAY: I'm generally against going down to subsubsection; not because it's complicated, but because it suggests that the organization of the paper isn't quite right. There should be a more linear/flat organization if the content is arranged right, imho. And then within the subsubsection you have \bold paragraph titles (which should be \paragraph{} in proper latex style). So maybe there are too many levels of hierarchy in this paper? Or maybe there's a simpler layout? There is also somewhat inconsistent use of \itemize, \enumerate, and these boldface paragraphs. So maybe take a look at the organization levels? My view is that the paper organization should be as hierarchical / complicated as the problem. And this problem is very simple.

\subsection{Data quality cuts as part of the selection function}\label{sec:quality_cuts}
If one wants to model a set of \emph{objects of interest}, say QSOs or WDs, it is often sensible to apply -- beyond the initial selection -- a number of 
% ARC drops: data quality cuts and cuts 
% to: 'cuts', since we are about to expand on that
cuts that eliminate spurious measurements, physical contaminants\footnote{By physical contaminants we mean astrophysical objects of a different nature than the objects of interest one seeks to mode, yet which project into nearly the same region of $\vq$ used for initial sample cuts}, and assure high data quality:\\

{\bf Selection function cuts to eliminate spurious measurements:\ }
It is often indispensable to eliminate objects from the parent catalog whose $\vq$ are not to be trusted -- commonly by means of data quality flags -- for two reasons: first, almost all modelling assumes that one understands the precision of the {\vq} in the catalog, i.e. the $\sigma_\vq$ must not be spurious. Second and related, the sample contamination from spurious objects, scattered to their seeming {\vq} as a result of poor attribute estimates, must be small; else it must be explicitly modelled. Such selection cuts, effectively terms in $S_\cat(\vq)$ that depend on a catalog's data quality flags, should foremost `clean' the sample; but one must check to which extent they affect the sample completeness, thereby altering $S_\cat(\vq)$. Such a check must be done empirically by applying an analogous selection cut on data quality flags to a sample known externally to be {\it bona fide} within the intended {\vq} regime. To the extent that data quality cuts do not affect the sample's completeness, they need not be treated in $S_\cat(\vq)$ in the subsequent modelling (Eq.~\ref{eq:SFdefinition}).

{\bf Cuts to reduce `physical contaminants':\ }  It is also often indispensable to apply additional cuts to the parent catalog, merely to separate the objects of interest from other classes of objects, `physical contaminants', with similar {\vq} used in the selection so far; such additional selection cuts also aim to boost the \emph{purity} of the sample. As above, such cuts ideally reduce only the number of contaminants, leaving the set of objects of interest untouched. In this limit, again the selection function would remain unchanged. As before, such cuts aim to make a simpler or better model for the remaining data; one may not need an elaborate model for the sample contaminants. The difference to the elimination of spurious objects is that the cuts to eliminate physical contaminants usually involve new and discriminating observables among the {\vq}, not data-quality flags. 
In good but realistic cases, such cuts can dramatically reduce the contamination while only eliminating a small fraction of the objects of interest. As above, the fraction of removed objects of interest must be determined as a function of the {\vq}, as they will lower $S_\cat(\vq)$. We will give a specific example of such cuts in Section~\ref{sec:WD-sample-selection}. \\\

{\bf Signal-to-Noise Cuts:} Signal-to-noise (S/N) cuts may be advisable for a number of reasons: often, the parent catalog has a vast set of catalog entries with marginal S/N in some attribute among the {\vq}. From a purely mathematical perspective, there is no cogent reason to eliminate such entries from consideration. Yet, modelling them requires an increasing, and often problematic, reliance (towards small S/N) on the precise and accurate estimate of catalog uncertainties. 
And it requires much more careful and explicit differentiation between the 'true' {\vq} and the 'cataloged' {\vq}. 
Given that $S_\cat(\vq)$ is a function of cataloged attributes, one must model which objects are being scattered in and out of the sample by $\vq_{obs}\ne \vq_{true}$(see \citealt{Frankel+2018,ED2020} and Section~\ref{sec:discussion}). 

In light of this, how should S/N cuts, such as $\plxSN > \plxSNlim$ be accounted for? In principle there are two options: We can include S/N attributes among the arguments of $S_\cat(\vq )$; but this then requires that the model  $\mathcal{M}(\vq\mid\modelpars)$ also predicts the uncertainties among the catalog attributes (e.g. predict both $\varpi$ and $\sigma_\varpi$). Or, almost equivalently, one can express the \emph{expected} S/N in term of other catalog `observables'. We focus on this latter approach in our worked example, where we make a cut in the \emph{expected} parallax S/N, $\eplxSN$, which we express as a function of {\mG} and $\varpi$; we do this because this straightforward approach may deserve wider use.

Taking  a selection by $\plxSN$  from the Gaia EDR3 catalog as an example, we now spell out the math of converting a S/N selection to one in terms of observables, as it does not seem to be documented in the literature.
We start with the simple scaling that reflects how $\plxSN$ varies with parallax and the ability to centroid a point source:
\begin{equation*}
\eplxSN \left(\varpi, \mG\right) \sim
\varpi~\cdot~ \sqrt{\operatorname{Flux}(\mG)}.
\end{equation*} 
One can then obtain a simple expression for the minimal parallax (maximal distance) at a given $\mG$ where 
the expected parallax S/N should exceed a threshold 
$\plxSNlim$:
\begin{equation}
    \varpi \ge \plxSNlim~\cdot~10^{\frac{\mG - \mG^{\rm r}}{5}}.
\label{eq:SNcut_appMag}
\end{equation}
The reference magnitude $\mG^{\rm r}$ in Eq.~\ref{eq:SNcut_appMag} can be derived from first principles, or scaled empirically: for Gaia EDR3 one finds $\mG^{\rm r}\approx 22$. 
Of course, it is likely, and in the case of Gaia known, that $\sigma_\varpi$, and by extension $\overline{S/N_\varpi}$ and $\mG^{\rm r}$, vary distinctly with position on the sky \citep{Lindegren2020,Everall2021}. In that case, the condition of Eq.~\ref{eq:SNcut_appMag} can simply become position-dependent if the level of accuracy is to be boosted.

%The same relation can be expressed as a function of \aMG~ and $\varpi$ as
%\begin{equation*}
%\eplxSN \bigl (\varpi, \aMG \bigr ) \sim
% \varpi^2~10^{\frac{-(\aMG - \mG^{\rm r}+10)}{5}}, 
%\end{equation*} 
%for parallaxes expressed in mas.

%\begin{equation}
%\varpi_{\rm min} \Bigl (\mG\mid \plxSNlim,m^{\rm r}_G\Bigr ) = \varpi \times %10^{({\mG-m^{\rm r}_G})/2.5},
%\label{eq:SNcut}
%\end{equation}
% with an empirical calibration of $m^{\rm r}_G\approx 21.0$ from EDR3.
% 
% {\color{red} I think the above eq. should be 
%\begin{equation}
%\varpi_{\rm min} \Bigl (\mG\mid \plxSNlim,m^{\rm r}_G\Bigr ) = \plxSNlim %\times 10^{({\mG-m^{\rm r}_G})/5},
%\end{equation}}

So, we can recapitulate the simple upshot of this example: when one aims to implement signal-to-noise selection criteria on one particular component of $\vq_i$, one could make the selection function an explicit function of measurement uncertainties (which would need to be modelled). However, it will often be preferable to keep the selection function simple (have fewer arguments) by instead applying selection cuts in observables among the {\vq} that amount to cuts in the \emph{expected} S/N in that $\vq_i$. But in general, sample cuts on S/N are perfectly legitimate, and often advantageous, if reflected correctly in $S_\cat(\vq )$. 

\subsection{Implementing the selection function in data - model comparisons}\label{sec:SF_in_data_model_comparison}

In its most general form, the role of the selection function in modelling is summarized in Eq.~\ref{eq:SFdefinition}, a simple multiplicative function of the {\vq}.  However, in many cases we want to compare only a few physical quantities between model and the data in the sub-sample, not the entire vector {\vq}. Indeed, quite often the selection function may depend on components of {\vq} that we do not want to model; then we need to marginalize out these `nuisance parameters'.

To make this concrete, we consider the example that we will work out in detail in the next Section: we want to learn about the luminosity-color function (LCF) of white dwarfs at the Sun's location in the Galaxy, $\Phi_0(M,c)$, which is the number density of WDs that have absolute magnitude and color $(M,c)$, a quantity with units $[\Mag^{-2}~\pc^{-3}]$. This density can of course vary with position, $\mathbf{x}$. We can write such a model, specified by parameters $\modelpars$, as
\begin{equation}
    \Phi(M,c,\mathbf{x}\mid\modelpars ) = \Phi_0(M,c\mid\modelpars) \, \ndenspars ,
    \label{eq:basic_model}
\end{equation}

where we have assumed that the spatial variation $\ndens$ separates multiplicatively. This may be convenient if our science interest is focussed on $\Phi_0(M,c)$ not on the dimensionless $\ndens $, which may be approximately known or be just less interesting; in almost all analogous modelling cases the actual positions of sources such as WDs $\{\mathbf{x}_i\}$ are of little astrophysical interest. Yet, they are crucial arguments in the selection function that reflects limits in sky coverage and distance (and implicitly the apparent magnitude). 
Therefore constraining $\Phi_0(M,c)$ requires us to marginalize over $\mathbf{x}$. The need for such marginalization is common, and must be  accounted for in the selection function.

Given a model that is phrased in terms of \emph{physical} quantities, here
$M,~c$ and 3D-position $\mathbf{x}$, Eq.~\ref{eq:SFdefinition} becomes
\begin{equation}
    \dd\Lambda_s (M,c,\mathbf{x}) = \Phi(M,c,\mathbf{x}\mid\modelpars)\, 
    S_\cat\bigl (\vq(M,c,\mathbf{x})\bigr )\  \dd M \dd c \  \dd V
    ~.
    \label{eq:dLambda}
\end{equation}
We now want to link this to the actual numbers of WDs with $(M,c)$ in our chosen sub-sample
(e.g. illustrated in Fig.~\ref{fig:sample-MC-distribution} below). For this we need to predict the expected number of entries  in the \emph{whole} sample (per $\dd M\cdot\dd c$) through volume-integration in Eq.~\ref{eq:dLambda}:
\begin{equation}
   \Lambda_\cat (M,c)= \Phi_0(M,c\mid\modelpars) \int \ndenspars \  S_\cat\bigl (\vq(M,c,\mathbf{x})\bigr )\  \dd^3\mathbf{x}\ ,
   \label{eq:Lambda}
\end{equation}
with $\Phi_0(M,c\mid\modelpars)$ and $\ndenspars$ from Eq.~\ref{eq:basic_model}; for compactness of notation we have dropped the explicit differentials $\dd \Lambda$ and $\dd M, \dd c$ for the remainder of the paper.

Note that if one wanted to constrain the spatial distribution of the tracers from the data at hand, rather than incorporate the presumed-to-be-known information about it in the modelling, we would of course not marginalize over the three spatial dimensions, but retain and model $\Lambda_s (M,c,\mathbf{x})$.

So, while the selection function appears initially as a multiplicative factor in the model prediction (Eq.~\ref{eq:SFdefinition}), in practice some or all of its dimensions are subject to marginalization integrals. This integration is over the quantities {\vq} that matter for the selection function but are mere nuisance parameters for the model.

\subsection{Is Sample Completeness Important?}
\label{sec:completeness}

It is often deemed the holy grail of sample design to be \emph{complete} with respect to some properties: `our parent sample contains basically all point sources in the sky with \mG\ brighter than $X$~magnitudes'; `our sample contains all stars of type $X$ within $Y$~parsec of the Sun', etc..
Sample completeness in this sense has an immediate visceral and intellectual appeal, with completeness to a magnitude limit or volume-completeness being perhaps the most common desiderata. 

But the merit of completeness in constraining models is much more nuanced: in many circumstances completeness is nice, but not necessary; in many other circumstances striving for completeness forces compromises in the sample design that lead to highly sub-optimal answers for the science questions of interest. 
We illustrate this crucial point here briefly and qualitatively, with a more quantitative underpinning in the worked WD example.

\noindent{\bf Completeness, nice but not necessary: } We know that the Gaia EDR3 catalog -- when averaged across the sky -- is nearly complete ($S\ge 0.95$) for magnitudes $12<\mG<19$ \citep{BE20,Smart20}. This is of course a fundamental piece of information to determine the incidence of astronomical phenomena and sources. But let us imagine we model a random 2/3 of the sources in the Gaia catalog, instead of the whole catalog. Then the sample is plenty large enough in most of the cases to do the model fitting; and if this 2/3 sub-sampling is properly reflected in the selection function and the modelling, we will get an identical modelling result. \emph{Knowing the level of completeness is far more important than being `complete'!} If there is no difference in the effort or resources needed to analyse the fully complete sample, there is no harm in it. But often -- think of spectroscopic follow-up of a photometric catalog --  striving for $S\equiv 1$ completeness implies an enormous additional resource effort; then the need for completeness must be justified by the necessity for answering the science question of interest.

\noindent{\bf `Complete' samples are often very sub-optimal: } Let us presume we want to estimate the space density of different objects (as in Section~\ref{M_c_WD}) that span a wide range in luminosities; and let's presume the common case that intrinsically faint objects are more numerous than more luminous objects. 
If we then want to construct a volume limited sample from some flux-limited parent catalog\footnote{`Volume-limited' in the sense of: nearly all objects of interest across all luminosities in this volume.} the maximal volume is set by the distance at which the least luminous objects fall below the basic flux limit of the parent catalog \citep[eg.][]{20pcSample2018,40pcSample2020}. The more luminous and more rare objects can be found in the parent catalog across much larger volumes. Yet, they get discarded for the sake of volume-completeness: 10-fold more luminous objects can be (naively) seen across $10^{3/2}\approx 30$-fold larger volumes, meaning that 97\% of them 
in the parent catalog get discarded from the volume complete sample.  If luminous objects are very rare, such stringent cuts to achieve volume-completeness may even leave them without any representation in the sample. As we will show in the next Section, there are selection function choices that are mathematically just as simple, provide unbiased model estimates, but can draw on much larger samples. And this situation is quite generic: \emph{Striving for (e.g. volume-) completeness just for its conceptual appeal may greatly increase the effort needed to get suitable data, or -- at a given data quality -- gravely limit the quality of the subsequent modelling!}

% \subsubsection{other aspects}
% touch on ever so briefly?
% \begin{enumerate}
%     \item selection on noisy quantities and error marginalisation
%     \item cross-catalog selection (including spectral selection)
%     \item {\tt repository for other aspects here}
% \end{enumerate}

\section{A Worked Example: the Color-Luminosity Function of White Dwarfs}\label{M_c_WD}

Following on these general considerations, we now turn to a worked example to flesh-out and illustrate the points above. For this we choose the \emph{luminosity-color function} (LCF) of White Dwarfs in the Galactic disk at $R_\odot$: their space density $\lcf$ as a function of absolute magnitude $M$ and color $c$. This is for a number of reasons: the example is of astrophysical interest; it can be implemented in a highly simplified form, with results differing (instructively) by orders of magnitude from the `naive' plotting of the face-value color-absolute magnitude diagram (CAMD); it allows us to illustrate almost all aspects from the above Section, and it can illustrate how much or how little impact sensible, but in detail arbitrary, choices in the sample selection make.

The distribution of WDs in mass-age space has long been recognized as a powerful diagnostic of both stellar physics and Galactic archaeology \citep{WD_Galactic_Archeology_Winget_1992,WD_Cosmochronology_Fontaine_2001}. Their mass distribution reflects the distribution of predecessor masses, in combination with the initial-to-final mass ratio \citep{WD_initial_to_final_mass_ratio_Weidemann_2000,El-Badry+IFMR}. Their distribution in luminosity at a given mass reflects both the birth-rate of such objects and their cooling histories. This is reflected in the \emph{observable} luminosity-color function, as those two observable quantities reflect a combination of WD mass and cooling age.

The spectacular data from Gaia DR2 \citep{GaiaDR2-HRD2018,Gentile-Fusillo+2019} immediately revealed how intricate the CAMD of WDs is in Gaia's \aMG\ vs.\ \BmR\ space:\footnote{Throughout this text we refer to the $(G_{\rm BP}-G_{\rm RP})$ color defined from the Gaia BP and RP bands as \BmR.} the WD distribution shows two branches in the CAMD around $\BmR\approx 0.2$, and the distribution shows a ridge across most colors at low luminosities, presumably related to the energy release following crystallization \citep{Cheng+2019,Tremblay2019}. This distribution can also be seen in Figure~\ref{fig:combined_Phi0} at the end of this paper. The detailed physical interpretation of this distribution's morphology is not yet settled \citep{Brown_Review_2021} and is beyond the scope of this paper.

Between the most luminous and dimmest WDs in typical Gaia-derived samples \citep[e.g.][]{Gentile-Fusillo+2019}  there are about 10 magnitudes, or a factor of 10,000 in luminosity. This implies that the volume across which WDs remain bright enough to enter a magnitude-limited sample varies by {\bf {\it many}} orders of magnitude\footnote{This situation is perfectly analogous to any galaxy survey, where luminous galaxies can be seen across vastly larger volumes than dwarf galaxies \citep[e.g.][]{Blanton2003}}. 

This vast range in effective WD survey volume must be accounted for. And, indeed has recently been done in determining the WD `luminosity function' \citep{Smart20}, which is the LCF integrated across the full color range at a given luminosity.
Yet, the intricacy and information-richness of the $\aMG\ vs. \ \BmR$ of WDs suggest that such an exercise should be generalized to retain the color-structure. We work this out here as an example: determining the WDs LCF, the space-density of WDs as a function of $\aMG$ and of color (where we take $\BmR$); subsequently we take $M$ and $c$ as shorthand for its two arguments. We will remain very cursory on the many astrophysical implications of this analysis in order to retain this paper's focus on the basics of how to devise and apply selection functions. 

\subsection{A Model for the Luminosity-Color Function of WDs}
The general LCF model has already been spelled out in Eq.~\ref{eq:basic_model}. But to actually make predictions, we  
need to specify functional forms for both the density normalization $\Phi_0(M,c\mid\modelpars)$ and for the dimensionless spatial density distribution $\ndenspars $. For $\Phi_0(M,c\mid\modelpars)$ we face the issue that there is no simple parameterized model that captures the white dwarfs CAMD patterns seen e.g. in Figure~\ref{fig:combined_Phi0}.  Therefore, we adopt a model where $\Phi_0(M,c\mid\modelpars)$ is described by independent top-hat functions within any small $(M,c)$ patch. If we then choose a $120\times 120$ grid in $(M,c)$, we have $14,400$ parameters $\modelpars$ for $\Phi_0(M,c\mid\modelpars)$.
For $\ndenspars $, which is essentially a nuisance parameter in the current context, we will adopt two very simplified functional forms, either a homogenous distribution, or a plane-parallel slab with a vertically Gaussian density profile of scale $h_z$. Note that we adopt a $\ndens$ here, and hence do not fit for parameters in $\ndenspars$; this is just one of the many `astrophysical choices' faced in model-building.

\subsection{WD Sample Selection}\label{sec:WD-sample-selection}
To constrain this model by confronting it with data, we need to choose a suitable sub-sample of WDs, which we now do. We start with an initial query to the Gaia EDR3 as our parent catalog, designed to yield an initial set of possible WD \emph{candidates} within a few hundred pc around the Sun with reasonably well-measured photometry and astrometry; we then refine this initial selection in a number of steps, resulting in an $S_\cat^{\rm sample}(\vq )$ that identifies WDs across their full parameter range with high purity.

Fundamentally, WDs can be  selected by their exceptional position in the CAMD, far below the main sequence at colors bluer than $\BmR\approx 2$. We want to capture WDs at all relevant colors (equivalently, surface temperatures), and in the end we want to eliminate most contaminants (i.e. sources that lie below the main sequence but are manifestly not single WDs). We also want to keep the sample as large as sensible  because our model for $\Phi_0(M,c)$  has many parameters. 

Following the considerations of the previous Section, we describe the sequence of selection cuts, which form multiplicative terms of the eventual sub-sample selection function  $S_\cat^{\rm sample}(\vq )$ that can be expressed solely as a function of $\vq=(\mG,\BmR,\varpi)$.

The initial Gaia EDR3 query encapsulates the following aspects:
\begin{itemize}
   \item We select on Gaia G band apparent magnitude $\mG < \mG^{\rm lim}\,\Mag$ with a fiducial magnitude limit of $\mG^{\rm lim}=20$, so that the approximation that Gaia EDR3 is `largely complete to that magnitude' is sensible.
    \item We choose the color range $-0.8 < \BmR < 2.5\,\Mag$, which entails basically the full color range expected for WDs. We take observed colors and ignore the issue of dust reddening for the time being.
    \item We apply a selection cut of parallax $\varpi > \varpi_{\rm lim} = 3\,\mas$, corresponding to a maximum sample extent of $333~$pc.
    While luminous WDs can be seen to greater distances, this choice eliminates the need for sophisticated models of the spatial density distribution of WDs, $\ndenspars $, and of sophisticated treatment of 3D dust extinction. To limit the size of the initial candidate WD sample, we also require {\tt parallax_over_error} $> 5$, a cut that will be superseeded by more stringent requirements in the subsequent analysis.
    \item We select objects that lie at least two magnitudes below the main sequence at that color through $\aMG(c) > M_{\rm MS}(c)+2$, with the absolute magnitude estimated as $\aMG \equiv \mG + 5\,\log_{10}\frac{\varpi}{100~\mas}$ (we design a sample where the difference between the true and the estimated absolute magnitude is negligible).
\end{itemize}
This translates into the following query to the Gaia EDR3 catalog:
\begin{lstlisting}[language=SQL]
SELECT *
FROM gaiaedr3.gaia_source 
WHERE
    phot_g_mean_mag < 20.0 
and bp_rp between -0.8 and 2.5
and parallax > 3.
and parallax_over_error > 5.
and phot_g_mean_mag +5*log10(parallax/100.) > 4.+ (13/3.3)*(bp_rp+0.8),
\end{lstlisting}
where the mean main sequence slope $\Delta \aMG / \Delta\BmR$ is adopted to be $13/3.3$.

\begin{figure}[ht]
    \centering
    \includegraphics[width=0.8\textwidth]{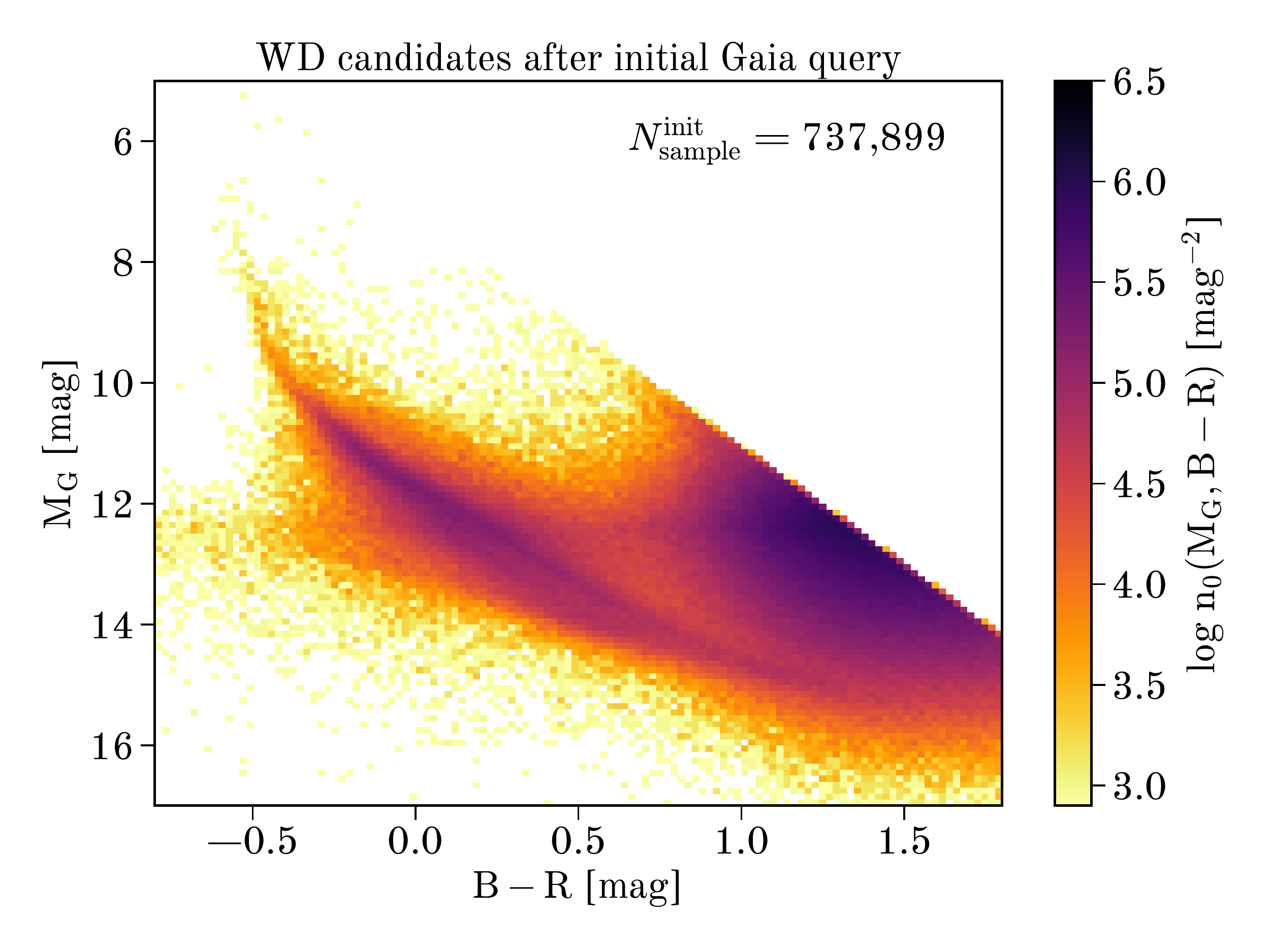}
    \caption{
    Distribution of objects in the color -- absolute magnitude diagram (CAMD, here (\BmR,\aMG)~) resulting from the initial sample query. This query selects WDs as it should -- the sequence of objects centered on $(\BmR, \aMG)=(0.2,12)$ --  but the initial candidate sample is overall dominated by objects seemingly near $(\BmR,\aMG)=(1.4,13)$; it turns out that almost all of these sources (seem to) lie in this portion of  the CAMD because they have spurious astrometry or photometry in EDR3 (see Fig.\ref{fig:high_astrometric_fidelity}). The units of this distribution is number of objects / mag$^2$.}
    \label{fig:initial_EDR3_query} 
\end{figure}

This query results in 737,899 returned Gaia EDR3 entries, whose CAMD is shown in Figure \ref{fig:initial_EDR3_query}, where we again equate $\aMG =  \mG + 5\,\log_{10}\frac{\varpi}{100~\mas}$. This Figure shows both the sequence of presumed WDs, around $(\BmR,\aMG)\approx (0.2,12)$, and a dominant set of other sources centered near $(\BmR,\aMG)\approx (1.4,13)$.
The diagonal right edge of the distribution reflects our well-below-the-main-sequence sample cut.
The latter group of sources, $(\BmR,\aMG)\approx (1.4,13)$,  turns out to be almost entirely spurious in their CAMD position, illustrating the need for sample cleaning. 

We now turn to additional cleaning cuts in the sample selection, made either to assure `data quality', or to achieve sufficient `astrophysical purity' (see Section~\ref{sec:quality_cuts}).

\begin{figure}[ht]
    \centering
    \includegraphics[width=0.8\textwidth]{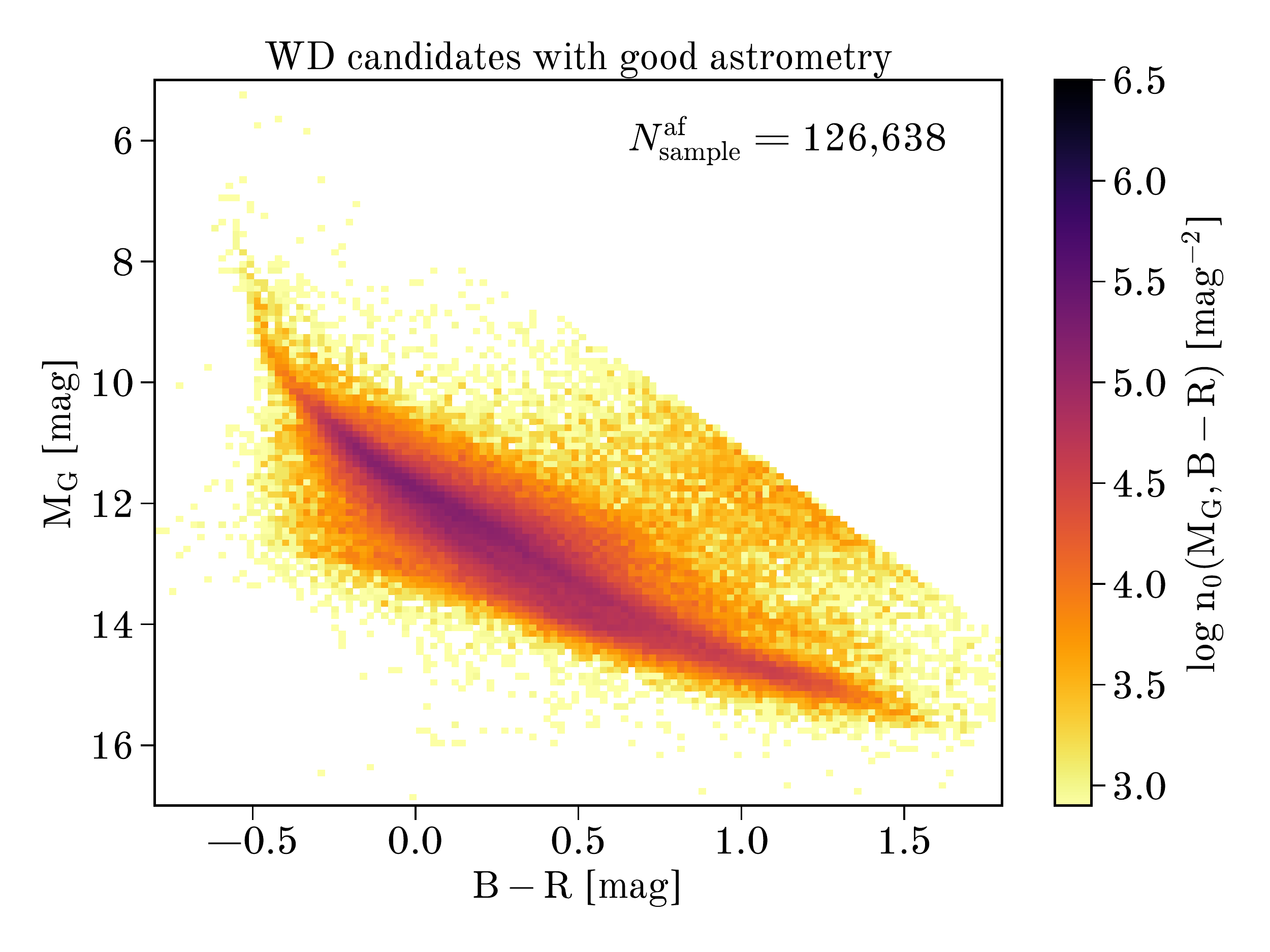}
    \caption{Distribution of objects in the CAMD after removing the (dominant subset of) presumably spurious objects, by requiring an astrometric fidelity of $\paf>0.9$ from \citet{Rybizki2021}. This cut, cast as $S^{\rm data~quality}_\cat(\paf\mid\vq)$, eliminates most spurious sources, but leaves $\ge 95\%$ of the spectroscopically confirmed WDs with $G<19.5$ unaffected.}
    \label{fig:high_astrometric_fidelity}
\end{figure}

\vspace{0.3cm}
\noindent$\bullet$~ We start with a 
     cut on the fidelity of the astrometric solution of the initial catalog entries. As Figure~\ref{fig:initial_EDR3_query} illustrates, objects that seemingly lie below the main sequence in a CAMD do so for chiefly two reasons: 
    their physical nature indeed places them there; or spurious astrometric (or color) measurements scatter them 
    there in rare instances. These rare instances, however, matter in this part of the CAMD, as it is extremely sparsely filled with {\it bona fide} objects: even rare spurious measurements may dominate the face-value population, as
    shown by \citet{Smart20} in this context. Therefore, a cut to remove spurious sources is indispensible here. We select on the astrometric fidelity parameter {\paf} recently proposed by
    \citet{Rybizki2021}, designed and verified to eliminate objects with spurious parallax measurements, specifically $\paf>0.9$. This removes the dominant fraction (83\%!) of all initial candidate sources as presumably spurious sources contaminating the sample. 
    The impact of this data quality cut is remarkable, as Figure~\ref{fig:high_astrometric_fidelity} shows: it removes almost all objects seemingly around $(\BmR,\aMG)\approx (1.4,13)$.  
    
    If this cut removed nothing but spurious objects, it would have {\it no} impact on the selection function. To check its impact on the objects of interest, we applied this cut to a sample of {\it bona fide} WDs \citep[spectroscopically verified by][]{WD_DR7},     
    and found that this quality cut only removed $\le 5\%$ of the WDs with $G<19.5$.
    In a more stringent analysis than executed here, one could and should empirically calibrate how this cut affects the selection function (as a function of $\mG$ and $\BmR$); below we simply neglect this few-percent effect. 
    Selection cuts on data-quality parameters such as these are generic to all analyses of large catalogs.

\begin{figure}
\centering 
\subfloat{%
  \includegraphics[width=0.45\textwidth]{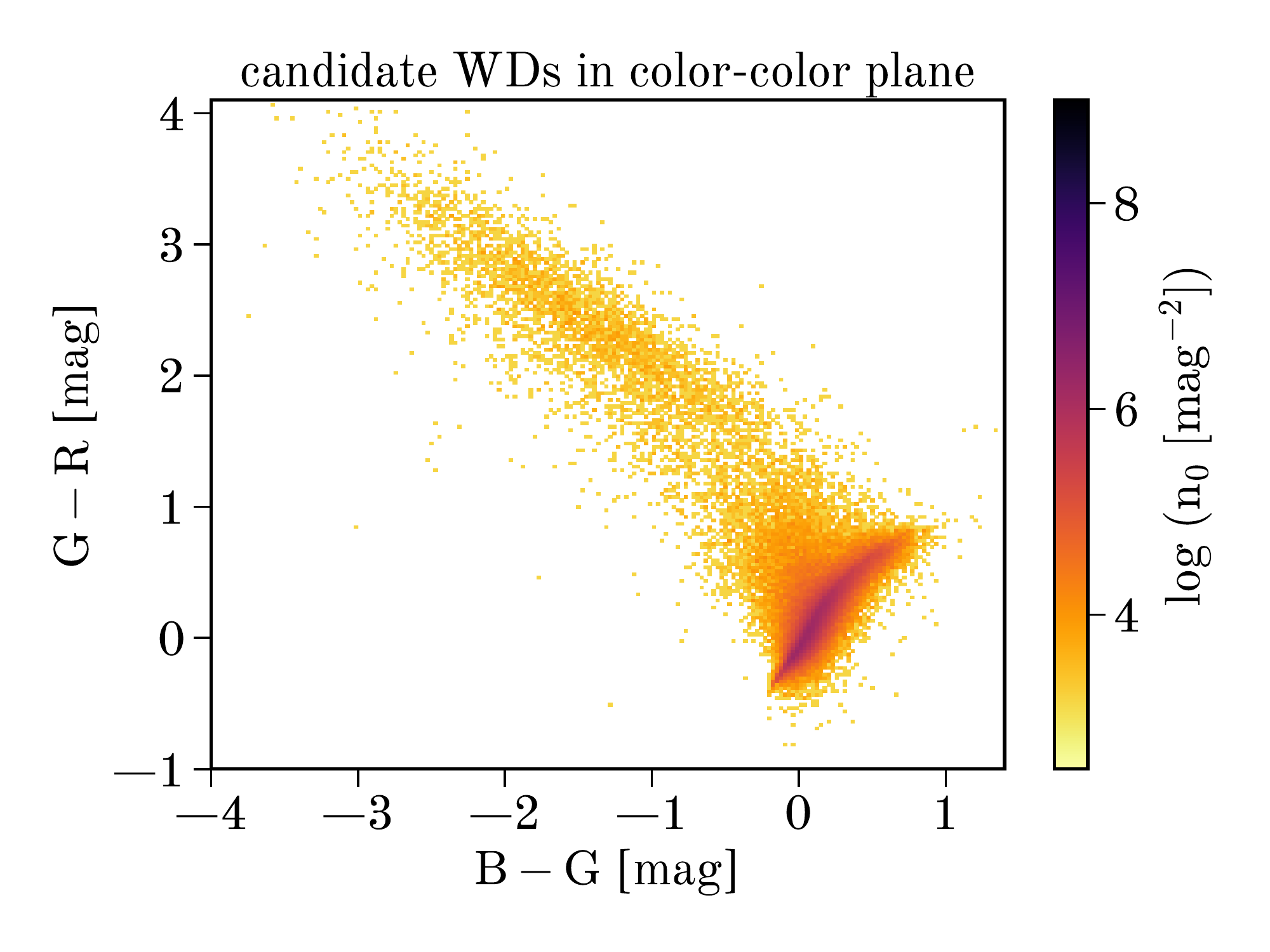}
  \label{fig:fig1}
}%\qquad
\subfloat{
  \includegraphics[width=0.45\textwidth]{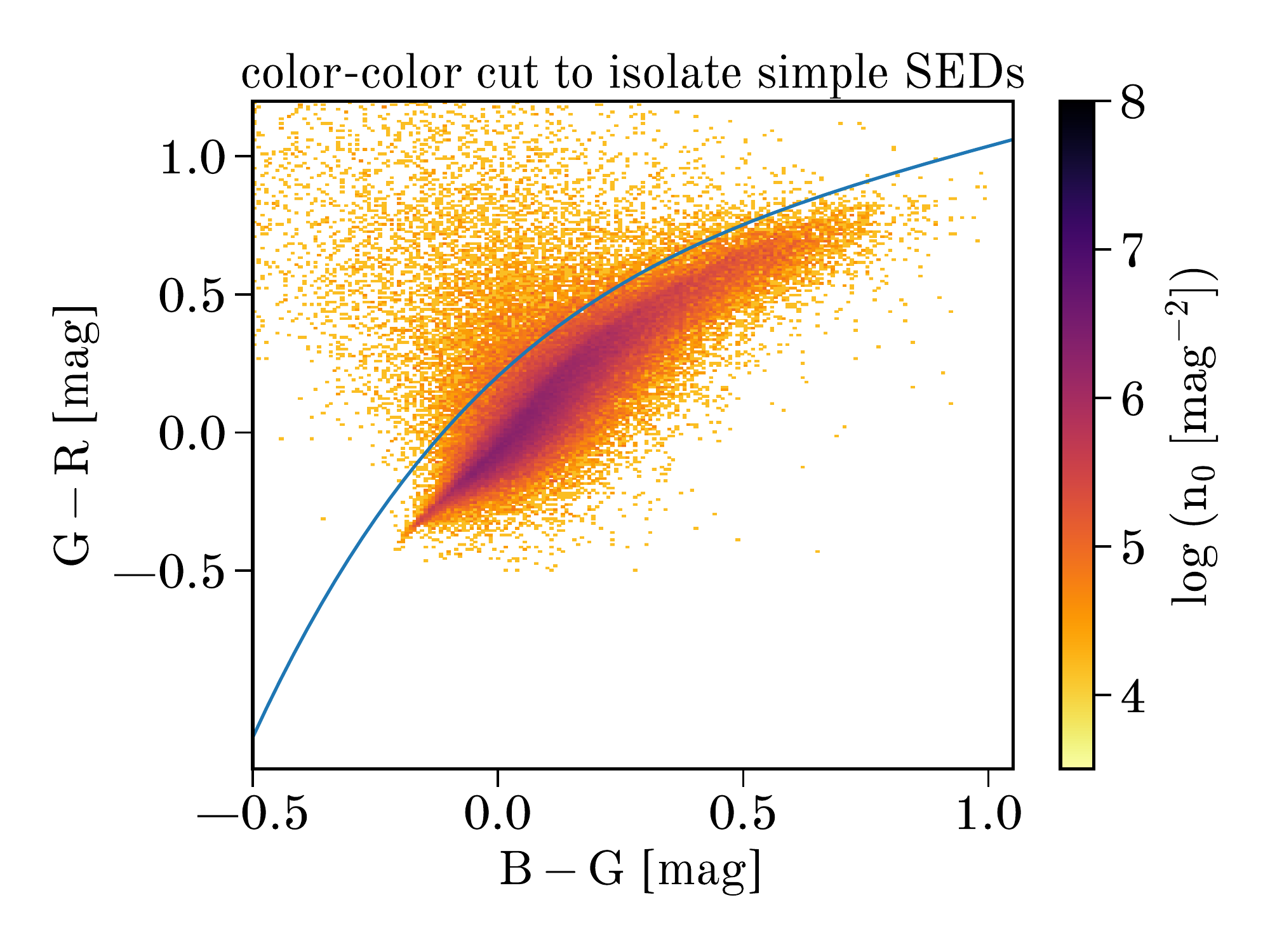}
  \label{fig:fig2}
}%
\caption{Distribution of the candidate WD sample (after the astrometric fidelity cut) in the 
  {\BmG} {\it vs.} {\GmR} color-color plane. The density distribution shows a sharp ridge where our objects of interest are located, objects with SEDs (or colors) of single WDs. The left panel, shows the full color distribution, which exhibits a subset with an enormous spread in colors: most are presumably binaries, involving a WD (possibly with an accretion disk) and a low-mass star. The color-cut, shown in the right panel as the blue line, eliminates most of those, while preserving $\ge 95\%$ of spectroscopically confirmed WDs. This is an example of a sample selection cut that leaves the selection function for the objects of interest essentially unaffected. It simply makes the sample purer, lessening the need to explicitly model contamination. }
  \label{fig:color-cuts}
\end{figure}

\begin{figure}[ht]
    \centering
    \includegraphics[width=0.8\textwidth]{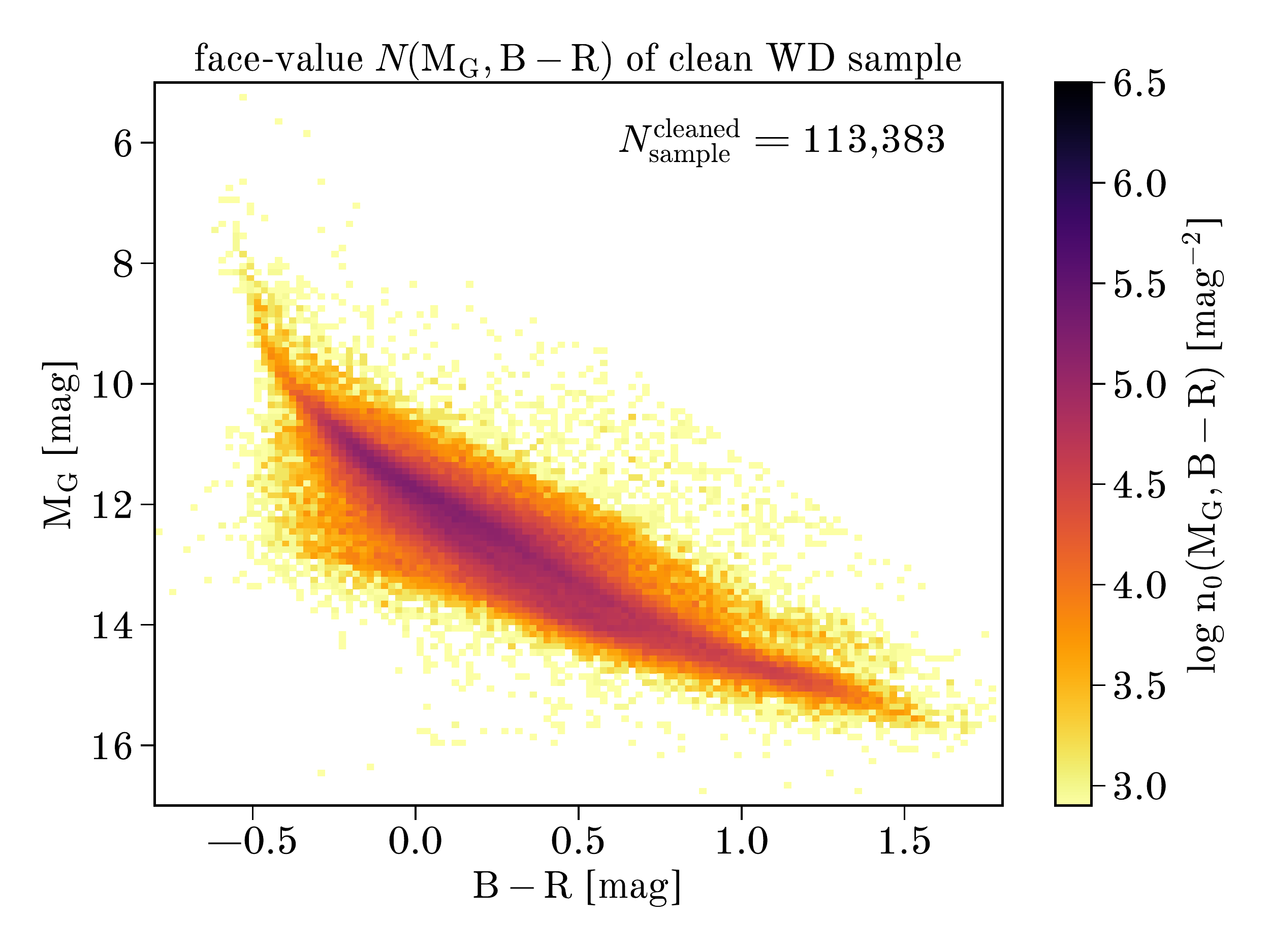}
    \caption{Face-value distribution of the WD sample members in the $N_\cat(\aMG,\BmR)$-plane, after all the selection function cuts summarized in Eq.\ref{eq:actualSF} have been applied. The units of this distribution is number of objects / mag$^2$.}
    \label{fig:sample-MC-distribution}
\end{figure}

\vspace{0.3cm}    
 \noindent$\bullet$ We now proceed to apply a cut that is designed to eliminate `physical' contaminants in this portion of the CAMD very effectively, yet leave the objects of interest (here WDs) essentially unaffected. The most common physical contaminants are presumably binary stars involving a WD and a low-mass main sequence star: these can be detached binaries, or in mass-transfer systems such as CVs, and are expected to be found in the vicinity of around $(\BmR,\aMG)\approx (1,12)$, where objects can be seen in \figurename~\ref{fig:high_astrometric_fidelity}.
 Single WDs (or two WDs of the similar $T_{\rm eff}$) form an extremely tight sequence in {\BmG} {\it vs.} {\GmR} space, as shown in Figure~\ref{fig:color-cuts}. Yet, almost all contaminating WD-MS binaries have an SED that is the combination of photospheres (or accretion disks) of very different $T_{\rm eff}$, scattering these objects over a wide area in the color-color locus (see left panel of Fig.\ref{fig:color-cuts}). To eliminate these objects, we apply a cut in the color plane, $\GmR < \GmR_{\rm lim}=f\left(\BmG\right)$, as indicated in the right panel of Figure~\ref{fig:color-cuts}:
    \begin{equation}
        \GmR < 0.48+1.15x-x^2+ 0.70x^3-0.2x^4\ \ \ \ \ x\equiv\BmG-0.2. 
        \label{eq:color-color-cut}
    \end{equation}
This cut eliminates $13,255$ sources (about 10\%).  Most of them are presumed physical contaminants (for WDs as objects of interest)
    as the resulting CAMD distibution shows (see Fig.\ref{fig:sample-MC-distribution}). This selection cut should not eliminate any of the objects of interest, as it encloses the full range $\BmG-\GmR$ expected for them. Only WDs with exceptionally poorly measured colors may get eliminated, and we checked that the impact of this cut on the spectroscopically verified WD sample from \citet{WD_DR7} was at the few percent level.
    
    While we will not model WD colors beyond (\BmR), any SED model for single WDs will predict zero probability for colors off the \emph{stellar locus} in the color-color plane; so the right-hand side of Eq.\ref{eq:SFdefinition} would stay unchanged for {\vq} away from this locus. An explicit treatment of such sample-purifying cuts may therefore not be necessary.
  
  \vspace{0.3cm}  
\noindent$\bullet$~ Finally, we turn to the issue of sub-selecting sources with `sufficiently precise parallax estimates'. As discussed in Section~\ref{sec:quality_cuts}, we do this through the condition 
  $\eplxSN (\varpi, \mG\bigr ) > \plxSNlim$, as per Eq.~\ref{eq:SNcut_appMag}. For estimating the WD LCF, we choose thresholds of 
  $\plxSNlim =5,10,{\mathbf{20}},40,80$. Expressing this S/N threshold through $\plxSNlim (\mG,\varpi)$ leaves us mostly (but not exclusively) with sample members whose directly determined $\plxSN$ is above the desired threshold; and we gain through this approach that we do not need to expand the arguments of the sub-sample selection function $S_\cat^{\rm sample}(\vq )$ beyond $\vq=\{\mG , \BmR , \varpi\}$.
  
Again, the value of any $\plxSNlim$ cut is an astrophysical \emph{choice} that can make uncertainties in \aMG\  less important, at the expense of reduced sample size.

To summarize, we create the sub-sample to be modelled from the Gaia EDR3 parent catalog through the following steps:

\noindent$\bullet${\bf Initial Sample Cuts:}  G$<G^{\rm lim}\approx 20$, $\varpi > \varpi_{\rm min}=3$~mas, $\varpi/\sigma_\varpi >5$,
              and sources 'below the main sequence': $\mG+5\log_{10}(\varpi/100) > 4 + 3.94\,(\BmR+0.8)$.\\
\noindent$\bullet$ {\bf Data Quality Cut}: $\paf>0.9$ from \citet{Rybizki+21} to eliminate spurious astrometry contaminants in this intrinsically sparse part of the CAMD.\\
\noindent$\bullet$ {\bf Physical Contaminant Elimination}: select only sources with SEDs resembling single WDs through a cut in the {\BmG} {\it vs.} {\GmR} plane (Figure~\ref{fig:color-cuts} and Eq.\ref{eq:color-color-cut}).\\
\noindent$\bullet$ {\bf Parallax S/N Cuts}: This is implemented through 
$\eplxSN (\varpi, \mG\bigr ) > \plxSNlim$, following 
Eq.\ref{eq:SNcut_appMag}, where we take $\plxSNlim = 20$ as fiducial, but explore other choices.\\

The full selection function, as used in Eq.~\ref{eq:SF_factorization}, consists of the following terms:
\begin{equation}
    \begin{aligned} 
        S_\cat(\vq) = &\, S_\cat^{\rm parent}(\vq) & \times \,
        S^{\rm init.~query}_\cat(\mG,\varpi,\sigma_\varpi,{\BmR}) \times
        S^{\rm data~quality}_\cat(\paf\mid\vq) \\
        & & \times\, S^{\rm contam.}_\cat(\BmG,\GmR) \times
        S^{\varpi SN}_\cat\bigl ( \eplxSN \mid \mG,\varpi\bigr )\,.
    \end{aligned}
   \label{eq:actualSF}
\end{equation}
We refer to each selection function term as a `cut', as we have implemented all terms in Eq.\ref{eq:SF_factorization} indeed as [0,1] step function reflecting Boolean conditions.

%\begin{figure}
%  \centering
%  \begin{subfigure}[b]{0.45\textwidth}
%    \centering
%    \includegraphics[width=.9\linewidth]{color_color_distribution1.pd%f}
%    %\caption{BP-G -- G-RP color distribution}
%    \label{fig:fig1}
%  \end{subfigure}%
%  \quad
%  \begin{subfigure}[b]{0.45\textwidth}
%    \centering
%    \includegraphics[width=.9\linewidth]{color_color_distribution2.pd%f}
%    %\caption{Color-cut in BP-G -- G-RP}
%    \label{fig:fig2}
%  \end{subfigure}
%  \caption{Distribution of the candidate WD sample (after the %astrometric fidelity cut) in the 
%  {\BmG} {\it vs.} {\GmR} color-color plane. The density distribution %shows a sharp ridge where our objects of interest are located, %objects with SEDs (or colors) of single WDs. The left panel, shows %the full color distribution, which exhibits a subset with an %enormous spread in colors: most are presumably binaries, involving %a WD (possibly with an accretion disk) and a low-mass star. The %color-cut, shown in the right panel as the blue line, eliminates %most of those, while preserving $\ge 95\%$ of spectroscopically %confirmed WDs. This is an example of a sample selection cut that %leaves the selection function for the objects of interest %essentially unaffected. It simply makes the sample purer, lessening %the need to explicitly model contamination. }
%  \label{fig:color-cuts}
%\end{figure}

These selection cuts lead to a catalog whose face-value number density distribution is shown in Figure~\ref{fig:sample-MC-distribution}, as a function of the key quantities $\left(M,c\right) =\left (\aMG,\BmR\right)$. It is worth iterating that all selection function cuts in Eq.~\ref{eq:actualSF} constitute examples of the types of cuts that many other analyses will also perform, whether they spell them out explicitly or not.

For ease in implementation of the subsequent modelling, we will make a number of further approximations whose astrophysical impact should be small.
First, we assume that the parent sample is complete across the sky, $S^{\rm parent}_\cat\approx 1\, \forall\, \mathbf{x}_{\rm sky}$. This is manifestly not true \citep[e.g.][]{BE20}, but at magnitudes brighter than $G\le 20$ it is a sky-averaged approximation that is good at the 5\% level. 
Next, we assume that we do not need to treat the terms 
$S^{\rm data~quality}_\cat(\paf\mid\vq)$ and $S^{\rm contam.}_\cat(\BmG,\GmR)$ explicitly, as these terms cut out almost exclusively spurious measurements and physical contaminants, but not our objects of interest, single WDs. Therefore, these terms are $\approx 1$ for all {\vq} where the model makes non-zero predictions. Again, this holds more broadly: selection cuts that only eliminate {\vq}-space where $\mathcal{M}(\modelpars)=0$ leave the selection function unaffected.

Finally, we will assume that the remaining terms of the selection function are only functions of three variables, $\vq=(\mG,\BmR,\varpi)$. Formally, the initial cut $S^{\rm init.~query}_\cat(\mG,\varpi,\sigma_\varpi,\BmR)$ is a function of  $\sigma_\varpi$, since it included the criteria $\frac{\varpi}{\sigma_\varpi} > 5$, but subsequent cuts in 
$\plxSNlim$ are more stringent, and expressed via $\vq=(\mG,\BmR,\varpi)$.

\subsection{Estimating $\Phi_0(M,c)$ from the Sub-Sample Defined by $S_\cat(\vq)$}\label{discussion}

% Also this step is very common in data modelling: the initial or face-value model prediction is linked to a catalog prediction via a marginalization integral over quantities that do not enter the data-comparison explicitly. For readability we omit the differentials $\dd\Lambda_{s}$ and $\dd M\dd c$ subsequently.
We now turn to working out the specifics of constraining our parameters, the $120\times 120$ elements of $\Phi_0(\aMG,\BmR)$, through the comparison with the data of this sub-sample,
$\{ \aMG (\mG,\varpi ),\BmR\}_{i=1,N_{\rm sample}}$. 
We start with the rate prediction for the catalog entries, $\Lambda_\cat $ from Eq.~\ref{eq:Lambda}. This prediction of $\Lambda_\cat (\aMG,\BmR)$ entails a marginalization over 3D space, as the detailed objects positions are not of interest here.
As often in astronomy, it is useful to separate such a volume integral in the angular components and the line-of-sight direction:
\begin{equation}
     \Lambda_\cat (\aMG,\BmR)= \Phi_0(\aMG,\BmR) \int_{4\pi}\dd\Omega\int_0^\infty d^2~ \dd d\  ~\ndenssky\ S_\cat\Bigl(\vq(\aMG,\mathbf{x}_{\rm sky},d)\Bigr ).
\end{equation}
From this we obtain with $\dd d/\dd\varpi = -\varpi^{-2}$
\begin{equation}
     \Lambda_\cat (\aMG,\BmR)= \Phi_0(\aMG,\BmR) \int_{4\pi}\dd\Omega\int_{\varpi_{\rm min}(\aMG|S_\cat(\vq ))}^\infty \varpi^{-4} \dd \varpi\ \hat{n}\bigl (\mathbf{x}_{\rm sky},d(\varpi)\bigr )
%    \ S_\cat\Bigl (\mG(\aMG,\varpi,\mathbf{x}_{\rm sky}),\varpi\Bigr )
     .
     \label{eq:rate-prediction}
\end{equation}
Note that in this case the impact of the selection function can be entirely subsumed in the lower bound of the last integral, $\varpi_{\rm min}(\aMG|S_\cat(\vq ))$; this is because we assumed that $S_\cat(\vq)\approx 1$ within these bounds, and zero beyond. Also, we can drop \BmR~ as an explicit argument of the selection function, as $S_\cat(\vq)$ does not vary significantly with color within the chosen range.

\begin{figure}[ht]
    \centering
    \includegraphics[width=0.8\textwidth]{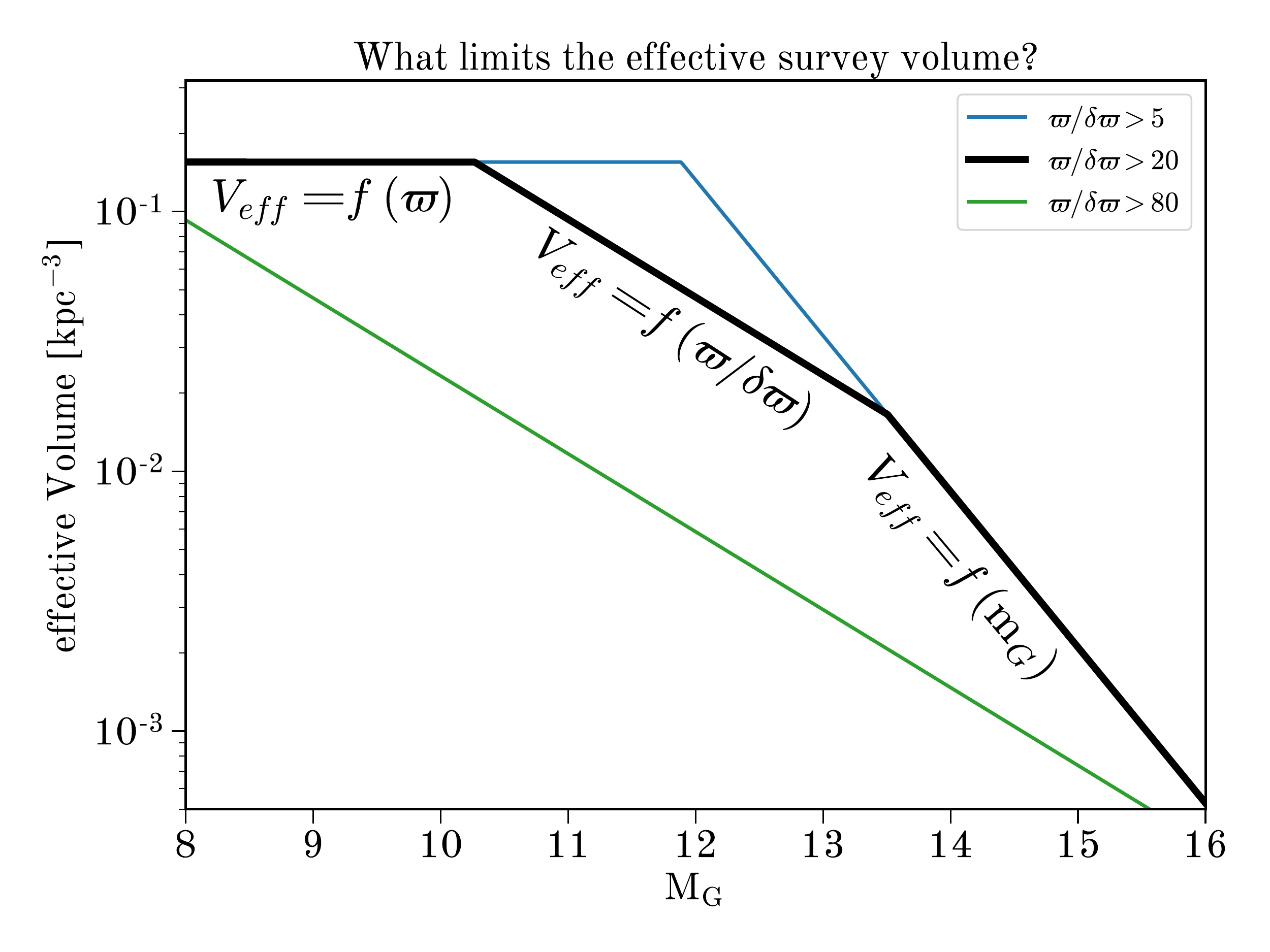}
    \caption{The effective survey volume, $\veff(\aMG)$, which  for our worked example is only a function of each object's estimated absolute magnitude $\aMG$. The Figure shows three regimes (thick black line) where different terms in $S_\cat(\vq)$ limit $\veff(\aMG)$: for the most luminous objects ($\aMG<11$) \veff\ is simply limited by the initial selection $\varpi>3$~mas; for the least luminous objects in the volume it is limited by the initial cut apparent magnitude, $\mG<20$. In the intermediate regime, the volume is limited by the (subsequently) chosen cut in \emph{expected} parallax S/N, $\left\langle\frac{\varpi}{\sigma_\varpi}\right\rangle_{\rm min}$ (Eq.\ref{eq:SNcut_absMag}). For very demanding choices in $\left\langle\frac{\varpi}{\sigma_\varpi}\right\rangle_{\rm min}$ this cut may dominate for all $\aMG$ (green line); if such a cut is omitted or very lenient (blue line), this regime may disappear. This Figure can also serve to illustrate why volume-limited samples are generally very sub-optimal: if we wanted to construct a volume-limited sample of WDs covering $7<\aMG<15$, it would have a volume of only $\veff=10^{-3}$~kpc$^{-3}$, and we would have to discard 90\% (99\%) of the accessible sample-members at $\aMG=13$ ($10$).}
    \label{fig:Veff}
\end{figure}

The formulation of Eq.~\ref{eq:rate-prediction} allows of course for angular variations in the selection function $S_\cat\bigl (\mG(\aMG,\varpi),\mathbf{x}_{\rm sky},\varpi\bigr )$ and for an arbitrary (but presumed known) density distribution $\hat{n}(\mathbf{x}_{\rm sky},d(\varpi))$. Here we approximate the entire selection function as isotropic, as its angular variation is orders of magnitude less important than its radial one (distance or parallax).

For this example and similar ones\footnote{The same holds whenever the model constrains `number densities' of objects of interest.} the integral of Eq.~\ref{eq:rate-prediction} boils down to calculating a properly weighted effective volume, \veff, because one can write the result as $\Lambda_\cat (\aMG,\BmR)= \Phi_0(\aMG,\BmR)\cdot \veff(\aMG)$. 

To evaluate this integral, we need to know how the non-zero domain of the selection function translates into the choice of the lower parallax integral boundary, $\varpi_{\rm min}$, in Eq.~\ref{eq:rate-prediction}. This limit varies of course among the sample members, as its depends on their $\mG$ and $\varpi$. The explicit selection function dependence is here only through $\varpi_{\rm min}=f\bigl ( \aMG(\mG,\varpi)\bigr )$. Considering the case at hand and neglecting dust extinction, the condition of assuring a minimal $\plxSNlim$ (Eq.~\ref{eq:SNcut_appMag}) translates into a minimal parallax of
\begin{equation}
    \varpi_{\rm min}\bigl ( \aMG\mid\plxSNlim\bigr )\equiv \sqrt{\,\plxSNlim}~\cdot~10^{\frac{\aMG - \mG^{\rm r}+10}{10}}.
\label{eq:SNcut_absMag}
\end{equation} 
The selection function specified in Eq.\ref{eq:actualSF} implies that the maximal distance $\varpi_{\rm min}$ can be set by three different aspects:
\begin{equation}
\varpi_{\rm min} \bigl (\aMG\mid S_\cat(\vq)\bigr ) = \max \begin{cases}
\varpi^{\rm lim},& \mathrm{initial}~\varpi~\mathrm{cut} \\
10^{(\aMG - \mG^{\rm lim}+10)/5}, & \mathrm{initial}~\mG~{\rm cut}\\
\sqrt{\,\plxSNlim} 10^{(\aMG - \mG^{\rm r}+10)/10},& \plxSNlim {\rm (Eq.~\ref{eq:SNcut_absMag})},
\end{cases}  
\label{eq:minparallax_cases}
\end{equation}
where $\varpi_{\rm min}$ is in units of mas, the absolute magnitude $\aMG(\mG,\varpi)$ is an implicit function of $\mG$ and $\varpi$, and $\mG^{\rm r}\approx 22 $ (for Gaia EDR3) is a reference magnitude for the sky-averaged parallax S/N scaling (see Eq.~\ref{eq:SNcut_appMag}). The simple form of Eq.~\ref{eq:minparallax_cases} is for the dust free case, but can get generalized to include dust exctinction (see Section~\ref{sec:discussion}).

This formulation also links this `forward modelling with a selection function' to classical \vmax-analyses \citep{Schmidt1968,Vmax_Avni_1980,Vmax_Paczynski_1990,Vmax_Lilly_1995}. The important difference here is that \veff\ is an effective volume that equals the true survey volume \vmax\ for objects of a given $\aMG$ only in the isotropic, homogeneous and dust-free limit.
Note that this sample has a well-defined {\veff} for each {\aMG}, but as an ensemble is \emph{not} volume limited in any global sense. As discussed in Section~\ref{sec:completeness}, a volume-limited sample 
with $\varpi_{\rm min} \bigl (\aMG\mid S_\cat(\vq)\bigr )= {\rm const.}$ for all $ \aMG$ would be sub-optimal and no simpler to model, as Eq.~\ref{eq:rate-prediction} with a variable $\varpi_{\rm min} \bigl (\aMG\mid S_\cat(\vq)\bigr )$ illustrates.

To proceed and actually evaluate Eq.~\ref{eq:rate-prediction} we now spell out two possible assumptions for the spatial density distribution of WDs: the first is that $\hat{n}\bigl (\mathbf{x}_{\rm sky},d(\varpi)\bigr )\approx{\rm const}$. This is clearly a poor approximation as soon as the sample reach becomes comparable or larger than the vertical scale height, $h_z$, of Galactic disk WDs. But it can prove instructive as the simplest limiting case. The second case, is to view the WD distribution near the Sun as a plane parallel slab of a single, and known (Gaussian) scale height $h_z$:
\begin{equation}
    \hat{n}\left(l,b,\varpi \right) = \exp\left(-\frac{z^2}{2h_z^2}\right) = \exp\left(- \frac{\sin^2{b}}{2\varpi^2 h_z^2}\right)\,
    \label{eq:spatial_density}
\end{equation}
Note that this neglects WDs density variations with Galactocentric radius and the likely age-dependence of the scale height, which has impact on {\veff} at the level of only a few percent. 

As stated before, we assume a) that, averaged over $4\pi$, dust extinction to the typical distance of the sample members is negligible; b) that the selection cuts in $\mG$ and $\varpi$ are sharp and uniform across the sky, allowing us to bypass the marginalization of the uncertainties in calculating \veff; and c) that the initial parent sample (and consequently the sub-samples) is approximately complete within these cuts. All these approximations may lead to systematic errors in estimating {\veff} that are, however, very small
compared to the \veff-range, $\veff(\aMG)$, with $\aMG$ ranging from 6 to 16~mag.

In general, the result of marginalizing over space in Eq.~\ref{eq:rate-prediction} can be expressed as
\begin{equation}
    \Lambda_\cat (\aMG,\BmR)= \Phi_0(\aMG,\BmR)\cdot \veff(\aMG)\,.
    \label{eq:Lambda_prediction}
\end{equation}

For the simplest, $\hat{n}\bigl (l,b,\varpi \bigr )=~$const., the effective volume {\veff} from Eq.~\ref{eq:rate-prediction} is analytic and intuitive, $\veff(\aMG)=\frac{4\pi}{3}\varpi_{\rm min}^{-3}(\aMG)$, where $\varpi_{\rm min}$ reflects the most stringent among three basic selection function choices as per Eq.~\ref{eq:minparallax_cases}: an initial parallax cut ($\varpi_{\rm lim}$), an initial apparent magnitude cut ($\mG$), and a subsequently chosen parallax S/N cut \plxSNlim.

For the more realistic case of $\hat{n}\bigl (l,b,\varpi \bigr )$ specified by Eq.\ref{eq:spatial_density}, the rate prediction still can be written compactly as in Eq.~\ref{eq:Lambda_prediction}, but $\veff(\aMG)$ is now generalized to 
\begin{equation}
    \veff(\aMG)= 2\pi\int_{-\pi /2}^{\pi / 2}\cos{b}~\dd b~\int_{\varpi_{\rm min}(\aMG)}^\infty \varpi^{-4} ~ \exp{\left(-\frac{\sin^2{b}}{2\varpi^2 h_z^2}\right)}~\dd \varpi \,.
    \label{eq:generalized_Veff}
\end{equation}

For population-averaged scale heights of $h_z\approx ~$300~pc \citep[e.g.][]{Bovy+2012a} and for $\varpi_{\rm min}$=3~mas the resulting {\veff} differ among the two approximations at the 15\% level for luminous WDs ($\aMG<10$), and only $\sim 1\%$ for the faintest WDs. We will nonetheless use the more accurate calculation of {\veff} from Eq.~\ref{eq:generalized_Veff} throughout the remainder of the analysis. In all density plots shown in the paper, this difference would not be discernable.

The resulting $\veff(\aMG)$ for our fiducial sample selection choices are shown in Figure~\ref{fig:Veff}, which indeed show that all three regimes of $\varpi_{\rm min}$ in Eq.~\ref{eq:minparallax_cases} come to bear in this regime. For the most luminous objects {\veff} is the same, set by the initial parallax cut $\varpi^{\rm lim}$; i.e. most luminous objects are volume complete. For the least luminous objects {\veff} is set by the maximal distance, or $\varpi_{\rm min}$, at which they
become fainter than $\mG^{\rm lim}$. For objects of intermediate luminosity, {\veff} is set by $\plxSNlim$. For a low threshold of $\plxSNlim$, say $>5$, this regime may be irrelevant; but for very demanding choices of $\plxSNlim$, say $>50$, this regime may dominate, as illustrated by the thin blue and green lines in Figure ~\ref{fig:Veff}.

We can now work out how to constrain the LCF at any one given color-pixel, $(\aMG,\BmR)$. This is eqivalent to 
asking what the probability of the model parameter $\Phi_0(\aMG,\BmR )$ is, given the number of sub-sample members in that pixel, $N_\cat(\aMG,\BmR )$ and $\veff(\aMG )$? For flat priors on $\Phi_0(\aMG,\BmR )$ and $N_\cat(\aMG,\BmR )$, this is just the Poisson probability  
\begin{equation}
    P\Bigl ( \Phi_0\mid N \Bigr ) = P\Bigl ( N\mid\Phi_0\Bigr )~\cdot ~\frac{ P(\Phi_0)}{P(N)} ~\sim ~\frac{\Lambda^N~e^{-\Lambda}}{N!},
    \label{eq:Possion_estimate_of_Phi0}
\end{equation}
where we use the shorthand of $\Phi_0$ for $\Phi_0(\aMG,\BmR )$, $N$ for $N_\cat(\aMG,\BmR )$, and $\Lambda$ for $\Lambda_\cat (\aMG,\BmR)\equiv \Phi_0(\aMG,\BmR )~\veff(\aMG )$.
If one is only interested in a simple point estimate for $\Phi_0(M,c)$ one can adopt
\begin{equation}
    \Phi_0(\aMG,\BmR) = \frac{N_\cat(\aMG,\BmR )}{\veff(\aMG )},
    \label{Phi0estimate}
\end{equation}
which is what we do for the plots in Section~\ref{sec:resulting_LCF}.

\begin{figure}[ht]
    \centering
    \includegraphics[width=0.8\textwidth]{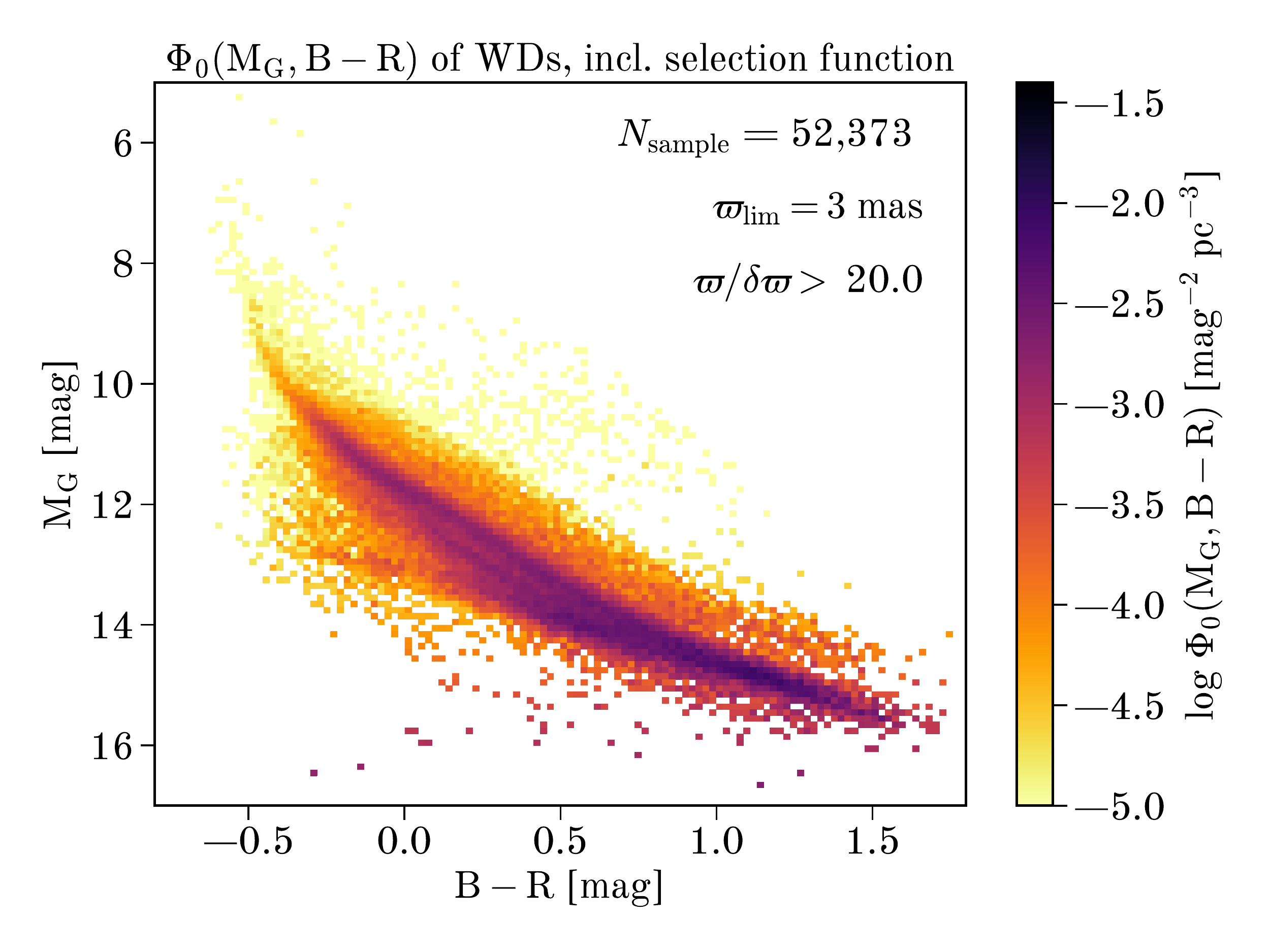}
    \caption{Estimate of $\Phi_0(\aMG,\BmR)$, the WD luminosity-color function (LCF), derived from Eq.~\ref{eq:Lambda_prediction} for the particular sample selection function choices listed within the Figure. Note that $\Phi_0(\aMG,\BmR)$ has units of [mag$^{-2}$~pc$^{-3}$].}
    \label{fiducial_Phi0}
\end{figure}

\begin{figure}
    \centering\includegraphics[width=4.5cm]{WD300pc_af_cc_uncorr.pdf}
    \centering\includegraphics[width=4.7cm]{WD300pc_plxSN_20.pdf}
    \centering\includegraphics[width=4.7cm]{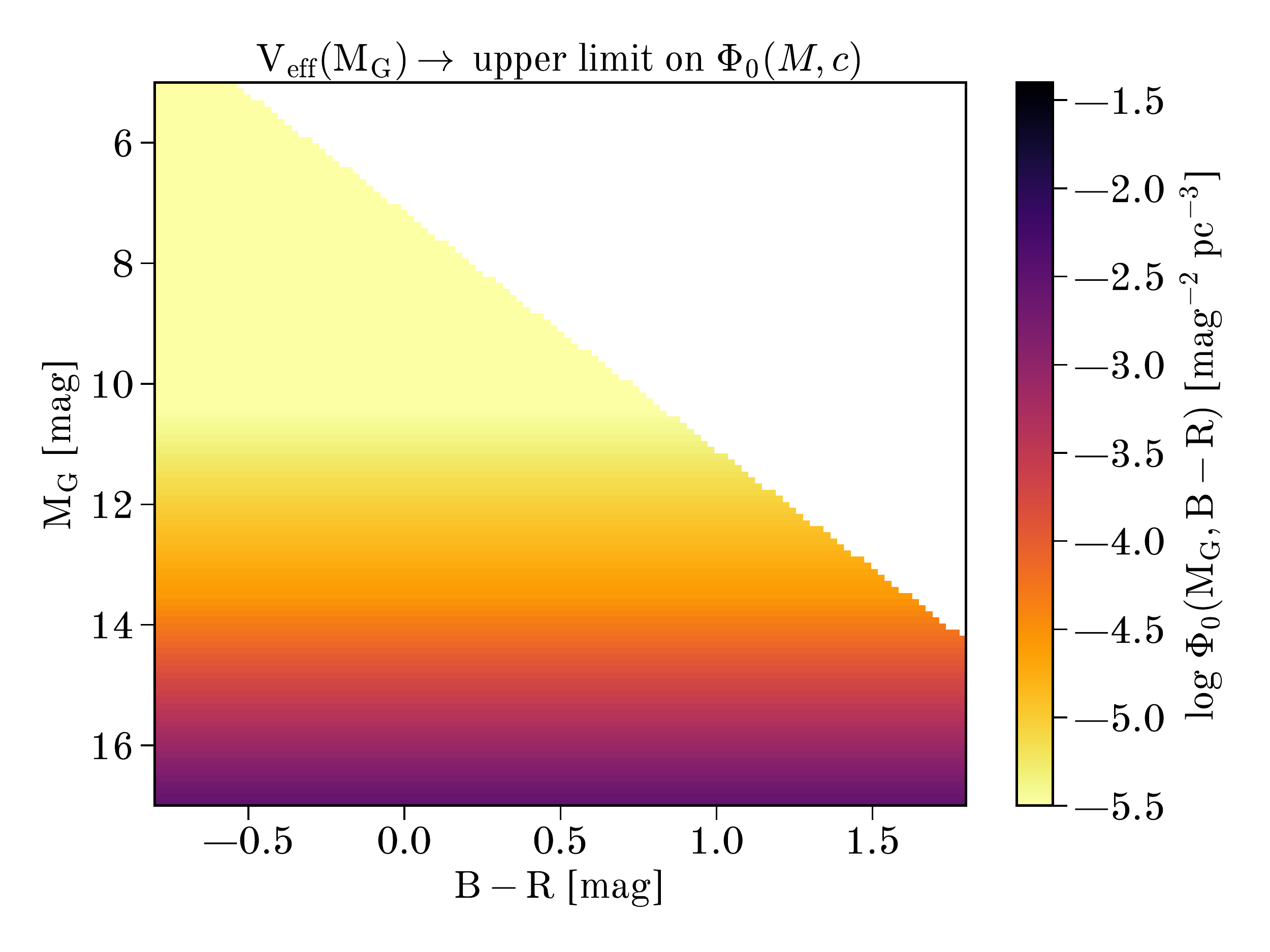}
  \caption{Comparison of the cleaned sample of likely WDs, $N_\cat(\aMG,\BmR)$~ (left, with units $[\mathrm{mag}^{-2}~]$) with  the modelled white dwarf LCF, $\Phi_0(\aMG,\BmR)$ ~(center, with units $[\mathrm{mag}^{-2}~\mathrm{pc}^{-3}]$). The (\aMG,\BmR) domain covered by both distribution is of course the same. Yet, the density peaks of $N_\cat(\aMG,\BmR)$ and of $\Phi_0(\aMG,\BmR)$ are in dramatically different places in the $(\aMG,\BmR)$-plane. The reason is that the two distributions differ by the selection function integral (expressed here via \veff), which is shown in the right panel; in the case at hand the selection function ends up being only a function of \aMG, not (\aMG,\BmR); the survey volume in the top right corner of the  $(\aMG,\BmR)$-plane is zero, as such objects were excluded by the initial Gaia parent catalog query.}
\label{Phi0-vs-Nsample}
\end{figure}

\begin{figure}
    \centering\includegraphics[width=\textwidth]{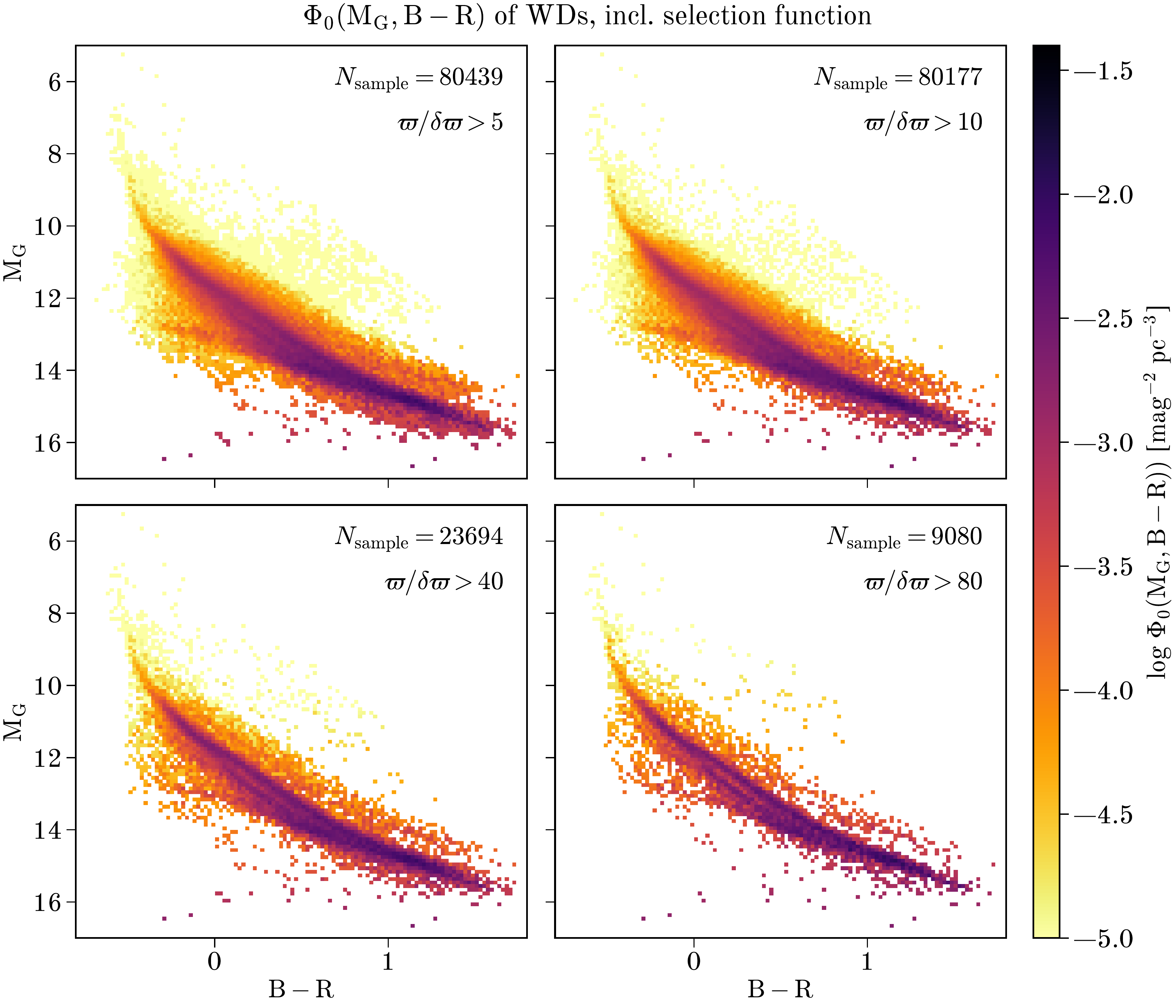}
 \caption{Estimate of $\Phi_0(\aMG,\BmR)$, analogous to that in Figure~\ref{fiducial_Phi0}, but for four different choices of $\plxSNlim$: 5,10,40,80. Despite the sample size differences of a factor of five, the inferred $\Phi_0(\aMG,\BmR)$ are mutually consistent. High values of $\plxSNlim$ lead to small samples but better detail in the high-density structure of $\Phi_0(\aMG,\BmR)$; low values of $\plxSNlim$ lead to larger samples and a better sampling of (\aMG,\BmR)-plane, but at reduced resolution, as the uncertainties in $\aMG$ become noticable.}  
\label{fig:plxSN_comparison}
\end{figure}

At first glance, the outcome of Eqs.~\ref{eq:rate-prediction}-\ref{eq:generalized_Veff} looks like a classic `volume correction approach'. But it is important to keep in mind that in the formulation of Eq.~\ref{eq:rate-prediction}, a number of the more subtle selection effects can be easily implemented, just requiring a numerical evaluation of the integral, as in the difference between Eq.~\ref{Phi0estimate} and Eq.~\ref{eq:generalized_Veff}; and this formulation will never apply\footnote{For convenience, we have phrased our estimate here as a volume correction in Eq.~\ref{Phi0estimate}; but of course, a probabilistic constraint on $\Phi_0(M,c)$ from
$p\left(N(M,c)\mid\Phi_0(M,c), \veff\right)$ would also work well.} any noise-amplifying  ``upward correction" of the data by division where $S_\cat\ll 1$.

\subsection{The Resulting Estimate of the White Dwarf LCF}\label{sec:resulting_LCF}

The basic point estimate of the WDs LCF is now simply an evaluation of Equation~\ref{Phi0estimate}, after choosing an $(\aMG,\BmR)$ grid on which our model parameters $\Phi_0(M,c)$ are to be evaluated. We know that the LCF has some fine-scale structure, and we therefore choose a fine grid of $120\times 120$ points in $\aMG$ and \BmR , covering $5<\aMG <17$ and $-0.8<\BmR <1.8$. Evaluation of Eq.\ref{Phi0estimate} results in the distribution shown in Figure~\ref{fiducial_Phi0}: the white dwarf LCF in the Galactic solar neighbourhood. Again, it is crucial to note that this density has units of $[\mathrm{mag}^{-2}~\mathrm{pc}^{-3}]$, the number of white dwarfs per magnitude-color interval and per volume.

Figure~\ref{Phi0-vs-Nsample} contrasts the face-value sample distribution of WDs $N_\cat(\aMG,\BmR )$ in the magnitude-color plane (left), with the estimate of the WDs LCF, $\Phi_0(\aMG,\BmR)$ (center). The two distributions cover the same domain in $(\aMG,\BmR)$, as per Eq.\ref{Phi0estimate}: the two panels differ only in being re-weighted line-by-line by the selection function through the corresponding {\veff} (right). But these quantitative differences are dramatic: while the sample-member density peaks near $(\aMG,\BmR)\approx (11,-0.1)$, the `true' LCF density peaks at $(\aMG,\BmR)\approx (15,1.2)$, where it is orders of magnitude higher than at $(11,-0.1)$. Of course, $N_\cat(\aMG,\BmR )$ and $\Phi_0(\aMG,\BmR)$ also differ by their units.

We believe that this quantitative LCF distribution of WDs, $\Phi_0(\aMG,\BmR)$,  deserves much astrophysical follow-up: e.g. testing WD evolutionary models, as it properly reflects the density along cooling tracks for different WD masses; or an estimate of the density of WDs along the crystalization line. Such analyses are beyond the scope of this paper, especially as they would benefit from the inclusion of spectroscopic WD classification information.

We now only turn to a few more technical points regarding the selection function. In particular, we want to focus on the impact of different $\plxSNlim$ choices on the analysis.  Figure~\ref{fig:plxSN_comparison} shows the same simple estimate of $\Phi_0(\aMG,\BmR)$ (Eq.\ref{Phi0estimate}) but for four alternate choices of \plxSNlim , namely $>5,10,40,80$; these different cuts lead to sample sizes that differ by nearly a factor of 10 in sample size. The first thing to note in Figure~\ref{fig:plxSN_comparison} is that the resulting density estimates of $\Phi_0(M,c)$ are mutually consistent, as they should be. An inclusive choice of $\plxSNlim = 5$ leads of course to a better sampled estimate of $\Phi_0(M,c)$. A far more stringent choice of $\plxSNlim >80$ leads to a far smaller sample, but one with very precise distances (and luminosities): this clarifies the bifurcation of
the LCF at intermediate colors and luminosities, yielding a sharper image of the LCF, but at the expense of sparser sampling.

\begin{figure}[ht]
    \centering
    \includegraphics[width=1.0\textwidth]{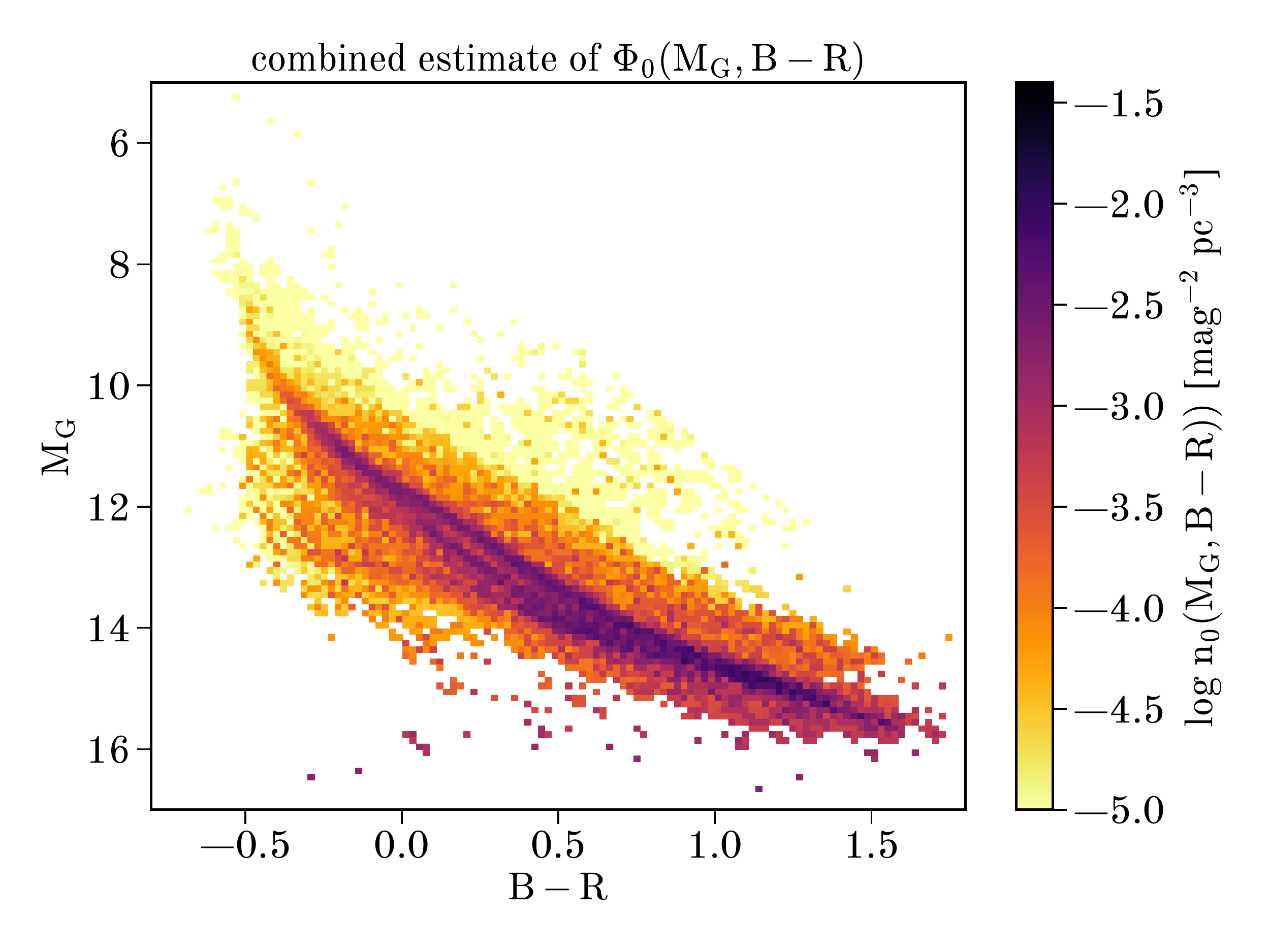}
    \caption{Estimate of the WD luminosity-color density (LCF) from a combination of the previous estimates. The high density regions were estimated using the $\plxSNlim >80$ sample, to take full advantage of the precise parallaxes for the sharp features in the diagram. The other parts were estimated from the $\plxSNlim >20$ sample, and finally from the $\plxSNlim >5$  sample that had been calculated on a coarser grid; these latter two steps reduce sampling noise in the low-density parts of the distribution. }
    \label{fig:combined_Phi0}
\end{figure}

This comparison also illustrates that the choices of sample cuts such as $\Delta\mathrm{S/N}(\varpi)$ are not universal, but should depend on the science goals. Of course, it is possible to combine the $\Phi_0(M,c)$ estimates resulting 
from different choices of \plxSNlim and possible different grid in $(\aMG,c)$. We show this in Figure~\ref{fig:combined_Phi0}: we started with the estimate resulting from $\plxSNlim >80$, which shows the sharpest high-density features but suffers from sparse sampling in the low-density regions. We retained the 20\% highest density pixels in that density map, replacing the rest with the values from the $\plxSNlim >20$ estimate; we repeated this exercise, retaining the 50\% highest density pixels in this $\Phi_0(M,c)$ estimate, and replaced the rest with the values from a LCD density map that had been constructed from a $2\times$ coarser $(\aMG,c)$ grid and a sample with $\plxSNlim >5$. The resulting LCF map (Fig.~\ref{fig:combined_Phi0}) combines sharp features in the LCF, including the bifurcation, with better S/N and coverage in the low-density regions. We did this in part to stress that the LCF, determined via Eq.~\ref{Phi0estimate} is a model estimate of a function defined across the full portion of the $(\aMG,c)$. 

\begin{figure}[ht]
    \centering
    \includegraphics[width=0.6\textwidth]{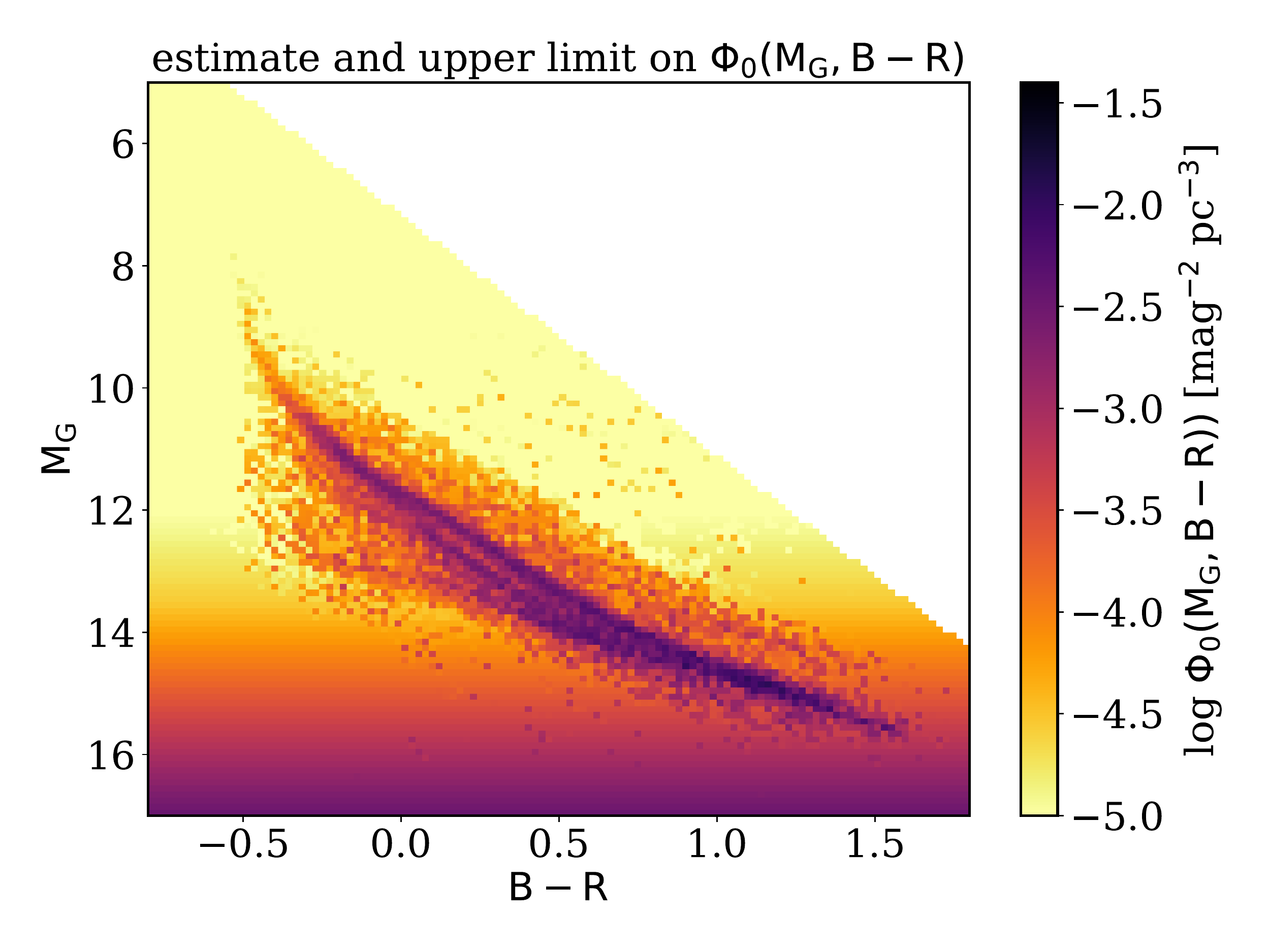}
    \caption{Representation of the $\Phi_0(\aMG,\BmR)$, as in Fig.~\ref{fig:combined_Phi0}, but aigmented by 1$\sigma$ upper limits across the entire (\aMG,\BmR) plane, where the sample does not contain any entries. Those limits were derived by calculating the density that even one single object in this (\aMG,\BmR)-pixel would imply. Knowing the selection function for counter-factual properties {\vq} permits to place constraints in models where we do not see any data in the sample.}
    \label{fig:upper_limits}
\end{figure}

This leads us to comment on the empty, white areas of the LCF distribution in the ($\aMG,\BmR$)-plane of, say, Figure~\ref{fig:combined_Phi0}. It is not that there are no constraints on the LCF for these ($\aMG,\BmR$), as the selection function is defined also for counter-factual objects. Indeed, 
the Poisson estimate for $\Phi_0(\aMG,\BmR)$ from Eq.~\ref{eq:Possion_estimate_of_Phi0} holds of course also for the empty pixels in the 
($\aMG,\BmR$)-plane, where $N_\cat(\aMG,\BmR)=0$. There, Eq.~\ref{eq:Possion_estimate_of_Phi0} implies e.g. a $1\sigma$ upper limit of
$\Phi_0(\aMG,\BmR) = 1/\veff(\aMG)$. So, this modelling implies estimates or upper limits across the entire ($\aMG,\BmR$) plane, as
illustrated in Figure~\ref{fig:upper_limits}.

These considerations lead to a compact way to present modelling results, as the ones presented here. The most immediate result is simply the 2D-array of $N_\cat(\aMG,\BmR)$, the most likely $\Phi_0(\aMG,\BmR)$ (Eq.~\ref{eq:Possion_estimate_of_Phi0})  along with 
the $(\aMG,\BmR)$ grid on which it is sampled. 
This allows to reconstruct the probability distribution for $\Phi_0(\aMG,\BmR)$.

%\section{Discussion and Summary}\label{sec:discussion_summary}
\section{Summary and Discussion}\label{sec:discussion}

We begin the final part of our selection function exposition by summarizing very briefly the selection function fundamentals. We then touch on a number of practical subtleties that need to be considered, especially in cases going beyond the worked example above and its simplifying assumptions.

\noindent$\bullet$\   \emph{What's a selection function? When do we need it?} In modelling `catalog data', a selection function is always needed when we want to ask questions about how frequently we are expected to find objects with certain physical attributes, their densities or property distribution; this type of modelling covers a very broad swath of astrophysical inquiry.  The selection function, $S_\cat(\vq )$, can be thought of as the multiplicative factor relating a model prediction, $\mathcal{M}(\vq\mid\modelpars)$ to a catalog incidence $ d\Lambda_\cat(\vq)$; or it can be thought of as the (dimensionless) probability that an object of properties {\vq} will be in a parent catalog, or a sub-sample drawn from it. 
    
\noindent$\bullet$\  \emph{How to Construct a Selection Function?} In practice, a selection function is constructed through a set of probabilistic conditions (often Boolean conditions) that describe the probability that an object (real or counter-factual) enters a (sub-)sample to be modelled. These conditions are intended to isolate near-optimal sets of objects that are informative about a physical question at hand. In general, there is no need or even benefit for samples to be `complete' with respect to any simple quantity such as flux or volume, just their selection function must be sufficiently well known.  Often, such sub-samples are drawn from a parent catalog (with its intrinsic selection function $S_\cat^{\rm parent}(\vq )$) and pared down to a suitable sub-sample by subsequent user-defined selection cuts, $S_\cat^{\rm sample}(\vq )$. And it usually makes sense to describe the overall selection function as the product of these two terms.  Ideally, the arguments of the selection function should be the minimal, or simplest, set of cataloged attributes {\vq} that 
    isolates a suitable sample, and that can be modelled (i.e. are arguments of $\mathcal{M}(\vq\mid\modelpars)$ ). In general that means that the {\vq} should be ``observables'' (positions, fluxes, etc..), but under many circumstances they can also be data-quality flags. 
    
\noindent$\bullet$\  \emph{How to Use Selection Functions in Modelling?} In most circumstances, there is no simple way to define or find an actually optimal selection function. The selection function $S_\cat(\vq)$ inevitably reflects astrophysically informed choices and judgements. However, it is crucial that any chosen selection function is properly applied in the subsequent modelling inference, where it is indispensable. In the simplest form, $S_\cat(\vq)$ appears in the modelling of its chosen sub-sample just as a multiplicative term. But in practice, $S_\cat(\vq)$ often depends -- for good reasons -- on quantities within {\vq} not used in the data-model comparison; these quantities are then best marginalized out (see Eq.~\ref{eq:Lambda})

Finally, we go through a number of complexities regarding the selection function that we have omitted so far in the interest of a simpler logical flow of the paper, as as they were not critical for our worked example.

\noindent$\bullet$\   \emph{Spatial complexity of selection functions:} The selection function of most large parent catalogs, $S_\cat^{\rm parent}$ has very complex spatial structure for a variety of reasons. For catalogs based on space-based all-sky surveys, the spacecraft's orbit and ensuing scanning pattern gives $S_\cat^{\rm parent}(m,\alpha,\delta)$ complex structure on the sky \citep[e.g.][]{BE20}. In addition crowding of sources can affect (lower) $S_\cat^{\rm parent}(m,\alpha,\delta)$ either because the overall source density is high, or if sources are spatially correlated, as for stellar clusters or binaries \citep[e.g.][]{El-Badry-2018}. In particular, the lower completeness to faint sources near bright sources is a selection function aspect that afflicts most survey catalogs.

\noindent$\bullet$\  \emph{Dust Extinction: Part of the Selection Function or of the Model?} Large catalogs that are based on X-ray, UV, optical or near-IR observations have observables that are affected by dust extinction. For catalogs of distant objects ($D\gtrsim 10$~kpc) dust is a foreground screen; for objects with 100~pc~$\lesssim D\lesssim$~10~kpc the 3D distribution of the dust matters \citep{DrimmelSpergel2001,Green+2019,Lallement2019}. This situation leaves several choices \citep[e.g.][]{Bovy+Dust2016}: To model observed or dereddened fluxes? To make extinction part of the selection function, $S_\cat(\vq )$, or part of the model, $\mathcal{M}(\vq\mid\modelpars)$? If the 2D/3D dust map was perfectly known, all these approaches are viable; they just differ in where the complexity is increased, either in the modelling or in the selection function. Yet two aspects argue for making dust extinction part of the modelling: It keeps the arguments {\vq} of the catalog incidence $\Lambda_\cat (\vq)$ and the selection function $S_\cat(\vq )$ better described as `observables'. And making the dust extinction part of the model allows us to marginalize out any uncertainties in it.

How dust extinction, say a 3D dust map $A_\lambda(l,b,\varpi )$, is implemented in the modelling depends on the details, though it always alters the model-predicted apparent magnitude $m(M,l,b,\varpi)$ and color, say, $\BmR(l,b,\varpi )$. In our worked example above, it just makes the maximal distance, $\varpi_{\rm min}$ to which an object is still in the sample a function of sky-position. The terms $\aMG-m_G^{lim/r}+10$ in the latter two cases of Eq.~\ref{eq:minparallax_cases} must be replaced by $\aMG-m_G^{lim/r}+10+A_G(l,b,\varpi )$; this makes Eq.~\ref{eq:minparallax_cases} an (easily solved) implicit equation, with a sky-position dependence through $A_G(l,b,\varpi )$. The resulting $\varpi_{\rm min}(l,b)$ 
then becomes an integration boundary that explicitly depends on sky position, $\varpi_{\rm min}(\mathbf{x}_{\rm sky},\aMG|S_\cat(\vq ))$ in the last integral of Eq.~\ref{eq:rate-prediction}:
\begin{equation}
     \Lambda_\cat (\aMG,\BmR)= \Phi_0(\aMG,\BmR) \int_{4\pi}\dd\Omega
     \int_{\varpi_{\rm min}(\mathbf{x}_{\rm sky},\aMG|S_\cat(\vq ))}^\infty \varpi^{-4} \dd \varpi\ \hat{n}\bigl (\mathbf{x}_{\rm sky},d(\varpi)\bigr ).
     \label{eq:dust-rate-prediction}
\end{equation}

For our worked example, the inclusion of 3D dust extinction would simply reduce the effective survey volume, $\veff(\aMG)$, which could be pre-computed for any given $\aMG$.  This integration again illustrates why the extinction model must exist also for counter-factual objects, i.e. all directions and distances that might be accessible in the dust-free case.
     
\noindent$\bullet$\  \emph{Sub-sample selection based on noisy catalog entries: } In the discussion so far, and in the worked example, we have assumed that the {\vq} in the catalog are precisely known at the values where $S_\cat(\vq )$ varies strongly (or, where $S_\cat(\vq )$ makes a cut). In that regime it is not necessary to differentiate explicitly between the `observed' $\vq_{obs}$ and the model-predicted $\vq_{true}$. However, in the majority of pertinent cases the uncertainties of the {\vq} at the selection function boundaries will matter: if one makes a sharp sample-selection cut at a certain $\vq_{obs}$, some sample members will have $\vq_{true}$ that lie beyond that selection boundary, while other objects that have $\vq_{true}$ within these boundaries will be absent from the chosen sub-sample. In that case one needs to have a model that covers a larger domain in {\vq} and needs to integrate over all $\vq_{true}$, accounting for the {\vq} uncertainties, $\sigma_\vq$.
In our notation,  Eq.~\ref{eq:Lambda} would generalize, for the
homoscedastic regime of a uniform, typical uncertainty, $\sigma_\vq$, to

\begin{equation}
    \dd\Lambda_s (M,c,\mathbf{x}) = \Phi(M,c,\mathbf{x}\mid\modelpars) 
    \int\dd\vq_{obs}~ S_\cat\bigl (\vq_{obs})\bigr )\ p\Bigl(\vq_{obs}\mid\vq_{true}(M,c,\mathbf{x}),\sigma_\vq\Bigr ) \dd M \dd c \  \dd V
    ~,
    \label{eq:dLambda_generalized}
\end{equation}
with corresponding changes in the equations that flow from Eq.~\ref{eq:Lambda}. Such an approach is applied explicitly e.g. in \citet{ForemanMackey-noisy-selection,Frankel+2018}.

\noindent$\bullet$\  \emph{What if the selection arises from the combination of two or more catalogs?} There are many instances, where the sub-sample selection arises from a combination of catalogs. The modelling of spectroscopic surveys, whose targets are almost inevitably drawn from a pre-existing photometric survey is a prime example. Modelling such spectroscopic surveys, say multi-object surveys such as SDSS, is based itself on a number of object selection steps: the selection of a sub-sample from photometric catalogs to yield objects eligible for spectroscopic targetting. This may be followed by the selection of objects among them that actually get spectroscopic observations, which then result in a new set of observables $\vq_{\rm spec}$. And ultimately one will select yet another sub-sample from among those, based on their $\vq_{\rm spec}$: e.g. selecting galaxies within a certain redshift range, stars of a certain metallicity, or stars spectroscopically classified as WDs. In most cases, these subsequent steps can still be treated as a sequence of multiplicative terms yielding $S_\cat(\vq)$, as in Eqs.~\ref{eq:SF_factorization}~\&~\ref{eq:actualSF}. Worked examples of this approach can be found e.g. in \citet{BR13} and \citet{Wojno2017}.

\noindent$\bullet$\   \emph{The precision of the selection function?} 
    In the discussion so far, we have presumed that the selection function has been determined with sufficient precision for whatever astrophysical problem is at hand. But we have not elaborated how one could assess the precision of $S_\cat(\vq ) = S^{\rm parent}_\cat(\vq )\cdot S^{\rm sample}_\cat(\vq )$, which will vary case-by-case. An example of how to assess the precision of  $S^{\rm parent}_\cat(\vq )$ is given in \citet{BE20}. In most cases, the precision of $S^{\rm sample}_\cat(\vq )$ should be high, as the selection functions are being designed in the same context as the modelling. However, there are cases where it is hard or even impossible to determine $S^{\rm sample}_\cat(\vq )$ in sufficient approximation, even if $S^{\rm parent}_\cat(\vq )$ is well-determined.
    The literature contains many samples of spectroscopic observations of objects contained in a catalog of well-defined $S^{\rm parent}_\cat(\vq )$. Let's consider the case where a selection function is not constructable in a quantitative way with respect to some quantity $a$ from among the {\vq}, while it is well understood with respect to another element of {\vq}, say $b$. As an example of such a situation we can take the radial metallicity distribution of stars in the Galactic disk from a spectroscopic survey, such as APOGEE \citep{APOGEE_2017}, $p([{\rm Fe}/{\rm H}],R)$. In this case, the selection function for the Galactocentric radii, $R(l,b,\varpi)$ may be hard to construct, but the spectroscopic targetting is insensitive to different [Fe/H] among disk stars \citep[see][]{Frankel+2018}.
    It would be difficult to model $p([{\rm Fe}/{\rm H}],R)$ for such a sample, but it is far easier to model the metalicity distribution \emph{conditioned} on Galactocentric radius, $p([{\rm Fe}/{\rm H}]\mid R)$. This can be generalized: if the selection function cannot be well determined for some components $(\vq_{\rm noSF}$ of $\vq)$ that one wants to model, one should build a model for the incidence of the other components of {\vq}, conditioned on $\vq_{\rm noSF}$.

None of the individual aspects of this selection function formalism is without precedence in the literature. But we hope that this paper can help to clarify when, and when not, all these aspects have to come together to do justice to the information content of large astronomical catalogs.

For this paper, we have restricted ourselves to applying this selection function formalism to the estimate of the luminosity-color function of white dwarfs. There are a number of other applications where exactly the same approach should be pursued, such as the LCF of cataclysmic variables, the LCF of the lowest main sequence into the brown dwarf regime, the LCF of hot subdwarfs, etc., as they all fall into the regime of nearby objects where the detailed spatial distribution is a nuisance parameter to be integrated out, but only under inclusion of an appropriate selection function.

The input files and the computations for all plots in this paper can be found in this \href{https://github.com/gaia-unlimited/WD-selection-function}{notebook}: https://github.com/gaia-unlimited/WD-selection-function.

\begin{acknowledgments}
The authors would like to thank Neige Frankel and Kareem El-Badry for thoughtful comments on the manuscript.

DB thanks Magdalen College for his fellowship and the Rudolf Peierls Centre for Theoretical Physics for providing office space and travel funds. AE thanks the Science and Technology Facilities Council of
the United Kingdom for financial support.

This work is a result from the GaiaUnlimited project which has received
funding from the European Union's Horizon 2020 research and innovation program
under grant agreement No 101004110. The GaiaUnlimited project was started at
the 2019 Santa Barbara Gaia Sprint, hosted by the Kavli Institute for
Theoretical Physics at the University of California, Santa Barbara.

This work has made use of data from the European Space Agency (ESA) mission
{\it Gaia} (\url{https://www.cosmos.esa.int/gaia}), processed by the {\it
Gaia} Data Processing and Analysis Consortium (DPAC,
\url{https://www.cosmos.esa.int/web/gaia/dpac/consortium}). Funding for the
DPAC has been provided by national institutions, in particular the
institutions participating in the {\it Gaia} Multilateral Agreement.
\end{acknowledgments}


\clearpage
\begin{thebibliography}{}
\expandafter\ifx\csname natexlab\endcsname\relax\def\natexlab#1{#1}\fi
\providecommand{\url}[1]{\href{#1}{#1}}
\providecommand{\dodoi}[1]{doi:~\href{http://doi.org/#1}{\nolinkurl{#1}}}
\providecommand{\doeprint}[1]{\href{http://ascl.net/#1}{\nolinkurl{http://ascl.net/#1}}}
\providecommand{\doarXiv}[1]{\href{https://arxiv.org/abs/#1}{\nolinkurl{https://arxiv.org/abs/#1}}}

\bibitem[{{Avni} \& {Bahcall}(1980)}]{Vmax_Avni_1980}
{Avni}, Y., \& {Bahcall}, J.~N. 1980, \apj, 235, 694, \dodoi{10.1086/157673}

\bibitem[{{Blanton} {et~al.}(2003){Blanton}, {Hogg}, {Bahcall}, {Brinkmann},
  {Britton}, {Connolly}, {Csabai}, {Fukugita}, {Loveday}, {Meiksin}, {Munn},
  {Nichol}, {Okamura}, {Quinn}, {Schneider}, {Shimasaku}, {Strauss}, {Tegmark},
  {Vogeley}, \& {Weinberg}}]{Blanton2003}
{Blanton}, M.~R., {Hogg}, D.~W., {Bahcall}, N.~A., {et~al.} 2003, \apj, 592,
  819, \dodoi{10.1086/375776}

\bibitem[{{Boubert} \& {Everall}(2020)}]{BE20}
{Boubert}, D., \& {Everall}, A. 2020, \mnras, 497, 4246,
  \dodoi{10.1093/mnras/staa2305}

\bibitem[{{Boubert} {et~al.}(2021){Boubert}, {Everall}, {Fraser}, {Gration}, \&
  {Holl}}]{Boubert21}
{Boubert}, D., {Everall}, A., {Fraser}, J., {Gration}, A., \& {Holl}, B. 2021,
  \mnras, 501, 2954, \dodoi{10.1093/mnras/staa3791}

\bibitem[{{Boubert} {et~al.}(2020){Boubert}, {Everall}, \& {Holl}}]{BEH20a}
{Boubert}, D., {Everall}, A., \& {Holl}, B. 2020, \mnras, 497, 1826,
  \dodoi{10.1093/mnras/staa2050}

\bibitem[{{Bovy}(2017)}]{Stellar_Inventory_Bovy_2017}
{Bovy}, J. 2017, \mnras, 470, 1360, \dodoi{10.1093/mnras/stx1277}

\bibitem[{{Bovy} \& {Rix}(2013)}]{BR13}
{Bovy}, J., \& {Rix}, H.-W. 2013, \apj, 779, 115,
  \dodoi{10.1088/0004-637X/779/2/115}

\bibitem[{{Bovy} {et~al.}(2016){Bovy}, {Rix}, {Green}, {Schlafly}, \&
  {Finkbeiner}}]{Bovy+Dust2016}
{Bovy}, J., {Rix}, H.-W., {Green}, G.~M., {Schlafly}, E.~F., \& {Finkbeiner},
  D.~P. 2016, \apj, 818, 130, \dodoi{10.3847/0004-637X/818/2/130}

\bibitem[{{Bovy} {et~al.}(2012){Bovy}, {Rix}, {Liu}, {Hogg}, {Beers}, \&
  {Lee}}]{Bovy+2012a}
{Bovy}, J., {Rix}, H.-W., {Liu}, C., {et~al.} 2012, \apj, 753, 148,
  \dodoi{10.1088/0004-637X/753/2/148}

\bibitem[{{Brown}(2021)}]{Brown_Review_2021}
{Brown}, A. G.~A. 2021, arXiv e-prints, arXiv:2102.11712.
\newblock \doarXiv{2102.11712}

\bibitem[{{Cheng} {et~al.}(2019){Cheng}, {Cummings}, \&
  {M{\'e}nard}}]{Cheng+2019}
{Cheng}, S., {Cummings}, J.~D., \& {M{\'e}nard}, B. 2019, \apj, 886, 100,
  \dodoi{10.3847/1538-4357/ab4989}

\bibitem[{{Cole} {et~al.}(2005){Cole}, {Percival}, {Peacock}, {Norberg},
  {Baugh}, {Frenk}, {Baldry}, {Bland-Hawthorn}, {Bridges}, {Cannon}, {Colless},
  {Collins}, {Couch}, {Cross}, {Dalton}, {Eke}, {De Propris}, {Driver},
  {Efstathiou}, {Ellis}, {Glazebrook}, {Jackson}, {Jenkins}, {Lahav}, {Lewis},
  {Lumsden}, {Maddox}, {Madgwick}, {Peterson}, {Sutherland}, \&
  {Taylor}}]{Cole2005}
{Cole}, S., {Percival}, W.~J., {Peacock}, J.~A., {et~al.} 2005, \mnras, 362,
  505, \dodoi{10.1111/j.1365-2966.2005.09318.x}

\bibitem[{{Drimmel} \& {Spergel}(2001)}]{DrimmelSpergel2001}
{Drimmel}, R., \& {Spergel}, D.~N. 2001, \apj, 556, 181, \dodoi{10.1086/321556}

\bibitem[{{El-Badry} \& {Rix}(2018)}]{El-Badry-2018}
{El-Badry}, K., \& {Rix}, H.-W. 2018, \mnras, 480, 4884,
  \dodoi{10.1093/mnras/sty2186}

\bibitem[{{El-Badry} {et~al.}(2018){El-Badry}, {Rix}, \&
  {Weisz}}]{El-Badry+IFMR}
{El-Badry}, K., {Rix}, H.-W., \& {Weisz}, D.~R. 2018, \apjl, 860, L17,
  \dodoi{10.3847/2041-8213/aaca9c}

\bibitem[{{Everall} {et~al.}(2021){Everall}, {Boubert}, {Koposov}, {Smith}, \&
  {Holl}}]{Everall2021}
{Everall}, A., {Boubert}, D., {Koposov}, S.~E., {Smith}, L., \& {Holl}, B.
  2021, \mnras, 502, 1908, \dodoi{10.1093/mnras/stab041}

\bibitem[{{Everall} \& {Das}(2020)}]{ED2020}
{Everall}, A., \& {Das}, P. 2020, \mnras, 493, 2042,
  \dodoi{10.1093/mnras/staa283}

\bibitem[{{Fontaine} {et~al.}(2001){Fontaine}, {Brassard}, \&
  {Bergeron}}]{WD_Cosmochronology_Fontaine_2001}
{Fontaine}, G., {Brassard}, P., \& {Bergeron}, P. 2001, \pasp, 113, 409,
  \dodoi{10.1086/319535}

\bibitem[{{Foreman-Mackey} {et~al.}(2014){Foreman-Mackey}, {Hogg}, \&
  {Morton}}]{ForemanMackey-noisy-selection}
{Foreman-Mackey}, D., {Hogg}, D.~W., \& {Morton}, T.~D. 2014, \apj, 795, 64,
  \dodoi{10.1088/0004-637X/795/1/64}

\bibitem[{{Frankel} {et~al.}(2018){Frankel}, {Rix}, {Ting}, {Ness}, \&
  {Hogg}}]{Frankel+2018}
{Frankel}, N., {Rix}, H.-W., {Ting}, Y.-S., {Ness}, M., \& {Hogg}, D.~W. 2018,
  \apj, 865, 96, \dodoi{10.3847/1538-4357/aadba5}

\bibitem[{{Gaia Collaboration} {et~al.}(2020{\natexlab{a}}){Gaia
  Collaboration}, {Brown}, {Vallenari}, {Prusti}, {de Bruijne}, {Babusiaux}, \&
  {Biermann}}]{eDR3}
{Gaia Collaboration}, {Brown}, A.~G.~A., {Vallenari}, A., {et~al.}
  2020{\natexlab{a}}, arXiv e-prints, arXiv:2012.01533.
\newblock \doarXiv{2012.01533}

\bibitem[{{Gaia Collaboration} \& et~al.(2016)}]{Gaia}
{Gaia Collaboration}, \& et~al. 2016, \aap, 595, A1,
  \dodoi{10.1051/0004-6361/201629272}

\bibitem[{{Gaia Collaboration} {et~al.}(2018){Gaia Collaboration}, {Babusiaux},
  {van Leeuwen}, {Barstow}, {Jordi}, {Vallenari}, {Bossini}, {Bressan},
  {Cantat-Gaudin}, {van Leeuwen}, {Brown}, {Prusti}, {de Bruijne},
  {Bailer-Jones}, {Biermann}, {Evans}, {Eyer}, {Jansen}, {Klioner}, {Lammers},
  {Lindegren}, {Luri}, {Mignard}, {Panem}, {Pourbaix}, {Randich}, {Sartoretti},
  {Siddiqui}, {Soubiran}, {Walton}, {Arenou}, {Bastian}, {Cropper}, {Drimmel},
  {Katz}, {Lattanzi}, {Bakker}, {Cacciari}, {Casta{\~n}eda}, {Chaoul}, {Cheek},
  {De Angeli}, {Fabricius}, {Guerra}, {Holl}, {Masana}, {Messineo}, {Mowlavi},
  {Nienartowicz}, {Panuzzo}, {Portell}, {Riello}, {Seabroke}, {Tanga},
  {Th{\'e}venin}, {Gracia-Abril}, {Comoretto}, {Garcia-Reinaldos}, {Teyssier},
  {Altmann}, {Andrae}, {Audard}, {Bellas-Velidis}, {Benson}, {Berthier},
  {Blomme}, {Burgess}, {Busso}, {Carry}, {Cellino}, {Clementini}, {Clotet},
  {Creevey}, {Davidson}, {De Ridder}, {Delchambre}, {Dell'Oro}, {Ducourant},
  {Fern{\'a}ndez-Hern{\'a}ndez}, {Fouesneau}, {Fr{\'e}mat}, {Galluccio},
  {Garc{\'\i}a-Torres}, {Gonz{\'a}lez-N{\'u}{\~n}ez}, {Gonz{\'a}lez-Vidal},
  {Gosset}, {Guy}, {Halbwachs}, {Hambly}, {Harrison}, {Hern{\'a}ndez},
  {Hestroffer}, {Hodgkin}, {Hutton}, {Jasniewicz}, {Jean-Antoine-Piccolo},
  {Jordan}, {Korn}, {Krone-Martins}, {Lanzafame}, {Lebzelter}, {L{\"o}ffler},
  {Manteiga}, {Marrese}, {Mart{\'\i}n-Fleitas}, {Moitinho}, {Mora}, {Muinonen},
  {Osinde}, {Pancino}, {Pauwels}, {Petit}, {Recio-Blanco}, {Richards},
  {Rimoldini}, {Robin}, {Sarro}, {Siopis}, {Smith}, {Sozzetti}, {S{\"u}veges},
  {Torra}, {van Reeven}, {Abbas}, {Abreu Aramburu}, {Accart}, {Aerts},
  {Altavilla}, {{\'A}lvarez}, {Alvarez}, {Alves}, {Anderson}, {Andrei},
  {Anglada Varela}, {Antiche}, {Antoja}, {Arcay}, {Astraatmadja}, {Bach},
  {Baker}, {Balaguer-N{\'u}{\~n}ez}, {Balm}, {Barache}, {Barata}, {Barbato},
  {Barblan}, {Barklem}, {Barrado}, {Barros}, {Bartholom{\'e} Mu{\~n}oz},
  {Bassilana}, {Becciani}, {Bellazzini}, {Berihuete}, {Bertone}, {Bianchi},
  {Bienaym{\'e}}, {Blanco-Cuaresma}, {Boch}, {Boeche}, {Bombrun}, {Borrachero},
  {Bouquillon}, {Bourda}, {Bragaglia}, {Bramante}, {Breddels}, {Brouillet},
  {Br{\"u}semeister}, {Brugaletta}, {Bucciarelli}, {Burlacu}, {Busonero},
  {Butkevich}, {Buzzi}, {Caffau}, {Cancelliere}, {Cannizzaro}, {Carballo},
  {Carlucci}, {Carrasco}, {Casamiquela}, {Castellani}, {Castro-Ginard},
  {Charlot}, {Chemin}, {Chiavassa}, {Cocozza}, {Costigan}, {Cowell}, {Crifo},
  {Crosta}, {Crowley}, {Cuypers}, {Dafonte}, {Damerdji}, {Dapergolas}, {David},
  {David}, {de Laverny}, {De Luise}, {De March}, {de Martino}, {de Souza}, {de
  Torres}, {Debosscher}, {del Pozo}, {Delbo}, {Delgado}, {Delgado}, {Diakite},
  {Diener}, {Distefano}, {Dolding}, {Drazinos}, {Dur{\'a}n}, {Edvardsson},
  {Enke}, {Eriksson}, {Esquej}, {Eynard Bontemps}, {Fabre}, {Fabrizio},
  {Faigler}, {Falc{\~a}o}, {Farr{\`a}s Casas}, {Federici}, {Fedorets},
  {Fernique}, {Figueras}, {Filippi}, {Findeisen}, {Fonti}, {Fraile}, {Fraser},
  {Fr{\'e}zouls}, {Gai}, {Galleti}, {Garabato}, {Garc{\'\i}a-Sedano},
  {Garofalo}, {Garralda}, {Gavel}, {Gavras}, {Gerssen}, {Geyer}, {Giacobbe},
  {Gilmore}, {Girona}, {Giuffrida}, {Glass}, {Gomes}, {Granvik}, {Gueguen},
  {Guerrier}, {Guiraud}, {Guti{\'e}}, {Haigron}, {Hatzidimitriou}, {Hauser},
  {Haywood}, {Heiter}, {Helmi}, {Heu}, {Hilger}, {Hobbs}, {Hofmann}, {Holland},
  {Huckle}, {Hypki}, {Icardi}, {Jan{\ss}en}, {Jevardat de Fombelle}, {Jonker},
  {Juh{\'a}sz}, {Julbe}, {Karampelas}, {Kewley}, {Klar}, {Kochoska}, {Kohley},
  {Kolenberg}, {Kontizas}, {Kontizas}, {Koposov}, {Kordopatis},
  {Kostrzewa-Rutkowska}, {Koubsky}, {Lambert}, {Lanza}, {Lasne}, {Lavigne}, {Le
  Fustec}, {Le Poncin-Lafitte}, {Lebreton}, {Leccia}, {Leclerc},
  {Lecoeur-Taibi}, {Lenhardt}, {Leroux}, {Liao}, {Licata}, {Lindstr{\o}m},
  {Lister}, {Livanou}, {Lobel}, {L{\'o}pez}, {Managau}, {Mann}, {Mantelet},
  {Marchal}, {Marchant}, {Marconi}, {Marinoni}, {Marschalk{\'o}}, {Marshall},
  {Martino}, {Marton}, {Mary}, {Massari}, {Matijevi{\v{c}}}, {Mazeh},
  {McMillan}, {Messina}, {Michalik}, {Millar}, {Molina}, {Molinaro},
  {Moln{\'a}r}, {Montegriffo}, {Mor}, {Morbidelli}, {Morel}, {Morris},
  {Mulone}, {Muraveva}, {Musella}, {Nelemans}, {Nicastro}, {Noval},
  {O'Mullane}, {Ord{\'e}novic}, {Ord{\'o}{\~n}ez-Blanco}, {Osborne}, {Pagani},
  {Pagano}, {Pailler}, {Palacin}, {Palaversa}, {Panahi}, {Pawlak},
  {Piersimoni}, {Pineau}, {Plachy}, {Plum}, {Poggio}, {Poujoulet},
  {Pr{\v{s}}a}, {Pulone}, {Racero}, {Ragaini}, {Rambaux}, {Ramos-Lerate},
  {Regibo}, {Reyl{\'e}}, {Riclet}, {Ripepi}, {Riva}, {Rivard}, {Rixon},
  {Roegiers}, {Roelens}, {Romero-G{\'o}mez}, {Rowell}, {Royer}, {Ruiz-Dern},
  {Sadowski}, {Sagrist{\`a} Sell{\'e}s}, {Sahlmann}, {Salgado}, {Salguero},
  {Sanna}, {Santana-Ros}, {Sarasso}, {Savietto}, {Schultheis}, {Sciacca},
  {Segol}, {Segovia}, {S{\'e}gransan}, {Shih}, {Siltala}, {Silva}, {Smart},
  {Smith}, {Solano}, {Solitro}, {Sordo}, {Soria Nieto}, {Souchay}, {Spagna},
  {Spoto}, {Stampa}, {Steele}, {Steidelm{\"u}ller}, {Stephenson}, {Stoev},
  {Suess}, {Surdej}, {Szabados}, {Szegedi-Elek}, {Tapiador}, {Taris}, {Tauran},
  {Taylor}, {Teixeira}, {Terrett}, {Teyssandier}, {Thuillot}, {Titarenko},
  {Torra Clotet}, {Turon}, {Ulla}, {Utrilla}, {Uzzi}, {Vaillant}, {Valentini},
  {Valette}, {van Elteren}, {Van Hemelryck}, {Vaschetto}, {Vecchiato},
  {Veljanoski}, {Viala}, {Vicente}, {Vogt}, {von Essen}, {Voss}, {Votruba},
  {Voutsinas}, {Walmsley}, {Weiler}, {Wertz}, {Wevers}, {Wyrzykowski},
  {Yoldas}, {{\v{Z}}erjal}, {Ziaeepour}, {Zorec}, {Zschocke}, {Zucker},
  {Zurbach}, \& {Zwitter}}]{GaiaDR2-HRD2018}
{Gaia Collaboration}, {Babusiaux}, C., {van Leeuwen}, F., {et~al.} 2018, \aap,
  616, A10, \dodoi{10.1051/0004-6361/201832843}

\bibitem[{{Gaia Collaboration} {et~al.}(2020{\natexlab{b}}){Gaia
  Collaboration}, {Smart}, {Sarro}, {Rybizki}, {Reyl{\'e}}, {Robin}, {Hambly},
  {Abbas}, {Barstow}, {de Bruijne}, {Bucciarelli}, {Carrasco}, {Cooper},
  {Hodgkin}, {Masana}, {Michalik}, {Sahlmann}, {Sozzetti}, {Brown},
  {Vallenari}, {Prusti}, {Babusiaux}, {Biermann}, {Creevey}, {Evans}, {Eyer},
  {Hutton}, {Jansen}, {Jordi}, {Klioner}, {Lammers}, {Lindegren}, {Luri},
  {Mignard}, {Panem}, {Pourbaix}, {Randich}, {Sartoretti}, {Soubiran},
  {Walton}, {Arenou}, {Bailer-Jones}, {Bastian}, {Cropper}, {Drimmel}, {Katz},
  {Lattanzi}, {van Leeuwen}, {Bakker}, {Casta{\~n}eda}, {De Angeli},
  {Ducourant}, {Fabricius}, {Fouesneau}, {Fr{\'e}mat}, {Guerra}, {Guerrier},
  {Guiraud}, {Jean-Antoine Piccolo}, {Messineo}, {Mowlavi}, {Nicolas},
  {Nienartowicz}, {Pailler}, {Panuzzo}, {Riclet}, {Roux}, {Seabroke}, {Sordo},
  {Tanga}, {Th{\'e}venin}, {Gracia-Abril}, {Portell}, {Teyssier}, {Altmann},
  {Andrae}, {Bellas-Velidis}, {Benson}, {Berthier}, {Blomme}, {Brugaletta},
  {Burgess}, {Busso}, {Carry}, {Cellino}, {Cheek}, {Clementini}, {Damerdji},
  {Davidson}, {Delchambre}, {Dell'Oro}, {Fern{\'a}ndez-Hern{\'a}ndez},
  {Galluccio}, {Garc{\'\i}a-Lario}, {Garcia-Reinaldos},
  {Gonz{\'a}lez-N{\'u}{\~n}ez}, {Gosset}, {Haigron}, {Halbwachs}, {Harrison},
  {Hatzidimitriou}, {Heiter}, {Hern{\'a}ndez}, {Hestroffer}, {Holl},
  {Jan{\ss}en}, {Jevardat de Fombelle}, {Jordan}, {Krone-Martins}, {Lanzafame},
  {L{\"o}ffler}, {Lorca}, {Manteiga}, {Marchal}, {Marrese}, {Moitinho}, {Mora},
  {Muinonen}, {Osborne}, {Pancino}, {Pauwels}, {Recio-Blanco}, {Richards},
  {Riello}, {Rimoldini}, {Roegiers}, {Siopis}, {Smith}, {Ulla}, {Utrilla}, {van
  Leeuwen}, {van Reeven}, {Abreu Aramburu}, {Accart}, {Aerts}, {Aguado},
  {Ajaj}, {Altavilla}, {{\'A}lvarez}, {{\'A}lvarez Cid-Fuentes}, {Alves},
  {Anderson}, {Anglada Varela}, {Antoja}, {Audard}, {Baines}, {Baker},
  {Balaguer-N{\'u}{\~n}ez}, {Balbinot}, {Balog}, {Barache}, {Barbato},
  {Barros}, {Bartolom{\'e}}, {Bassilana}, {Bauchet}, {Baudesson-Stella},
  {Becciani}, {Bellazzini}, {Bernet}, {Bertone}, {Bianchi}, {Blanco-Cuaresma},
  {Boch}, {Bombrun}, {Bossini}, {Bouquillon}, {Bragaglia}, {Bramante},
  {Breedt}, {Bressan}, {Brouillet}, {Burlacu}, {Busonero}, {Butkevich},
  {Buzzi}, {Caffau}, {Cancelliere}, {C{\'a}novas}, {Cantat-Gaudin}, {Carballo},
  {Carlucci}, {Carnerero}, {Casamiquela}, {Castellani}, {Castro-Ginard},
  {Castro Sampol}, {Chaoul}, {Charlot}, {Chemin}, {Chiavassa}, {Cioni},
  {Comoretto}, {Cornez}, {Cowell}, {Crifo}, {Crosta}, {Crowley}, {Dafonte},
  {Dapergolas}, {David}, {David}, {de Laverny}, {De Luise}, {De March}, {De
  Ridder}, {de Souza}, {de Teodoro}, {de Torres}, {del Peloso}, {del Pozo},
  {Delgado}, {Delgado}, {Delisle}, {Di Matteo}, {Diakite}, {Diener},
  {Distefano}, {Dolding}, {Eappachen}, {Edvardsson}, {Enke}, {Esquej}, {Fabre},
  {Fabrizio}, {Faigler}, {Fedorets}, {Fernique}, {Fienga}, {Figueras},
  {Fouron}, {Fragkoudi}, {Fraile}, {Franke}, {Gai}, {Garabato},
  {Garcia-Gutierrez}, {Garc{\'\i}a-Torres}, {Garofalo}, {Gavras}, {Gerlach},
  {Geyer}, {Giacobbe}, {Gilmore}, {Girona}, {Giuffrida}, {Gomel}, {Gomez},
  {Gonzalez-Santamaria}, {Gonz{\'a}lez-Vidal}, {Granvik},
  {Guti{\'e}rrez-S{\'a}nchez}, {Guy}, {Hauser}, {Haywood}, {Helmi}, {Hidalgo},
  {Hilger}, {H{\l}adczuk}, {Hobbs}, {Holland}, {Huckle}, {Jasniewicz},
  {Jonker}, {Juaristi Campillo}, {Julbe}, {Karbevska}, {Kervella}, {Khanna},
  {Kochoska}, {Kontizas}, {Kordopatis}, {Korn}, {Kostrzewa-Rutkowska},
  {Kruszy{\'n}ska}, {Lambert}, {Lanza}, {Lasne}, {Le Campion}, {Le Fustec},
  {Lebreton}, {Lebzelter}, {Leccia}, {Leclerc}, {Lecoeur-Taibi}, {Liao},
  {Licata}, {Lindstr{\o}m}, {Lister}, {Livanou}, {Lobel}, {Madrero Pardo},
  {Managau}, {Mann}, {Marchant}, {Marconi}, {Marcos Santos}, {Marinoni},
  {Marocco}, {Marshall}, {Polo}, {Mart{\'\i}n-Fleitas}, {Masip}, {Massari},
  {Mastrobuono-Battisti}, {Mazeh}, {McMillan}, {Messina}, {Millar}, {Mints},
  {Molina}, {Molinaro}, {Moln{\'a}r}, {Montegriffo}, {Mor}, {Morbidelli},
  {Morel}, {Morris}, {Mulone}, {Munoz}, {Muraveva}, {Murphy}, {Musella},
  {Noval}, {Ord{\'e}novic}, {Orr{\`u}}, {Osinde}, {Pagani}, {Pagano},
  {Palaversa}, {Palicio}, {Panahi}, {Pawlak}, {Pe{\~n}alosa Esteller},
  {Penttil{\"a}}, {Piersimoni}, {Pineau}, {Plachy}, {Plum}, {Poggio},
  {Poretti}, {Poujoulet}, {Pr{\v{s}}a}, {Pulone}, {Racero}, {Ragaini},
  {Rainer}, {Raiteri}, {Rambaux}, {Ramos}, {Ramos-Lerate}, {Re Fiorentin},
  {Regibo}, {Ripepi}, {Riva}, {Rixon}, {Robichon}, {Robin}, {Roelens},
  {Rohrbasser}, {Romero-G{\'o}mez}, {Rowell}, {Royer}, {Rybicki}, {Sadowski},
  {Sagrist{\`a} Sell{\'e}s}, {Salgado}, {Salguero}, {Samaras}, {Sanchez
  Gimenez}, {Sanna}, {Santove{\~n}a}, {Sarasso}, {Schultheis}, {Sciacca},
  {Segol}, {Segovia}, {S{\'e}gransan}, {Semeux}, {Shahaf}, {Siddiqui},
  {Siebert}, {Siltala}, {Slezak}, {Solano}, {Solitro}, {Souami}, {Souchay},
  {Spagna}, {Spoto}, {Steele}, {Steidelm{\"u}ller}, {Stephenson},
  {S{\"u}veges}, {Szabados}, {Szegedi-Elek}, {Taris}, {Tauran}, {Taylor},
  {Teixeira}, {Thuillot}, {Tonello}, {Torra}, {Torra}, {Turon}, {Unger},
  {Vaillant}, {van Dillen}, {Vanel}, {Vecchiato}, {Viala}, {Vicente},
  {Voutsinas}, {Weiler}, {Wevers}, {Wyrzykowski}, {Yoldas}, {Yvard}, {Zhao},
  {Zorec}, {Zucker}, {Zurbach}, \& {Zwitter}}]{Smart20}
{Gaia Collaboration}, {Smart}, R.~L., {Sarro}, L.~M., {et~al.}
  2020{\natexlab{b}}, arXiv e-prints, arXiv:2012.02061.
\newblock \doarXiv{2012.02061}

\bibitem[{{Gentile Fusillo} {et~al.}(2019){Gentile Fusillo}, {Tremblay},
  {G{\"a}nsicke}, {Manser}, {Cunningham}, {Cukanovaite}, {Hollands}, {Marsh},
  {Raddi}, {Jordan}, {Toonen}, {Geier}, {Barstow}, \&
  {Cummings}}]{Gentile-Fusillo+2019}
{Gentile Fusillo}, N.~P., {Tremblay}, P.-E., {G{\"a}nsicke}, B.~T., {et~al.}
  2019, \mnras, 482, 4570, \dodoi{10.1093/mnras/sty3016}

\bibitem[{{Green} {et~al.}(2019){Green}, {Schlafly}, {Zucker}, {Speagle}, \&
  {Finkbeiner}}]{Green+2019}
{Green}, G.~M., {Schlafly}, E., {Zucker}, C., {Speagle}, J.~S., \&
  {Finkbeiner}, D. 2019, \apj, 887, 93, \dodoi{10.3847/1538-4357/ab5362}

\bibitem[{{Hollands} {et~al.}(2018){Hollands}, {Tremblay}, {G{\"a}nsicke},
  {Gentile-Fusillo}, \& {Toonen}}]{20pcSample2018}
{Hollands}, M.~A., {Tremblay}, P.~E., {G{\"a}nsicke}, B.~T., {Gentile-Fusillo},
  N.~P., \& {Toonen}, S. 2018, \mnras, 480, 3942, \dodoi{10.1093/mnras/sty2057}

\bibitem[{{Kleinman} {et~al.}(2013){Kleinman}, {Kepler}, {Koester}, {Pelisoli},
  {Pe{\c{c}}anha}, {Nitta}, {Costa}, {Krzesinski}, {Dufour}, {Lachapelle},
  {Bergeron}, {Yip}, {Harris}, {Eisenstein}, {Althaus}, \&
  {C{\'o}rsico}}]{WD_DR7}
{Kleinman}, S.~J., {Kepler}, S.~O., {Koester}, D., {et~al.} 2013, \apjs, 204,
  5, \dodoi{10.1088/0067-0049/204/1/5}

\bibitem[{{Lallement} {et~al.}(2019){Lallement}, {Babusiaux}, {Vergely},
  {Katz}, {Arenou}, {Valette}, {Hottier}, \& {Capitanio}}]{Lallement2019}
{Lallement}, R., {Babusiaux}, C., {Vergely}, J.~L., {et~al.} 2019, \aap, 625,
  A135, \dodoi{10.1051/0004-6361/201834695}

\bibitem[{{Lilly} {et~al.}(1995){Lilly}, {Tresse}, {Hammer}, {Crampton}, \& {Le
  Fevre}}]{Vmax_Lilly_1995}
{Lilly}, S.~J., {Tresse}, L., {Hammer}, F., {Crampton}, D., \& {Le Fevre}, O.
  1995, \apj, 455, 108, \dodoi{10.1086/176560}

\bibitem[{{Lindegren} {et~al.}(2020){Lindegren}, {Klioner}, {Hern{\'a}ndez},
  {Bombrun}, {Ramos-Lerate}, {Steidelm{\"u}ller}, {Bastian}, {Biermann}, {de
  Torres}, {Gerlach}, {Geyer}, {Hilger}, {Hobbs}, {Lammers}, {McMillan},
  {Stephenson}, {Casta{\~n}eda}, {Davidson}, {Fabricius}, {Gracia-Abril},
  {Portell}, {Rowell}, {Teyssier}, {Torra}, {Bartolom{\'e}}, {Clotet},
  {Garralda}, {Gonz{\'a}lez-Vidal}, {Torra}, {Abbas}, {Altmann}, {Anglada
  Varela}, {Balaguer-N{\'u}{\~n}ez}, {Balog}, {Barache}, {Becciani}, {Bernet},
  {Bertone}, {Bianchi}, {Bouquillon}, {Brown}, {Bucciarelli}, {Busonero},
  {Butkevich}, {Buzzi}, {Cancelliere}, {Carlucci}, {Charlot}, {Cioni},
  {Crosta}, {Crowley}, {del Peloso}, {del Pozo}, {Drimmel}, {Esquej}, {Fienga},
  {Fraile}, {Gai}, {Garcia-Reinaldos}, {Guerra}, {Hambly}, {Hauser},
  {Jan{\ss}en}, {Jordan}, {Kostrzewa-Rutkowska}, {Lattanzi}, {Liao}, {Licata},
  {Lister}, {L{\"o}ffler}, {Marchant}, {Masip}, {Mignard}, {Mints}, {Molina},
  {Mora}, {Morbidelli}, {Murphy}, {Pagani}, {Panuzzo}, {Pe{\~n}alosa Esteller},
  {Poggio}, {Re Fiorentin}, {Riva}, {Sagrist{\`a} Sell{\'e}s}, {Sanchez
  Gimenez}, {Sarasso}, {Sciacca}, {Siddiqui}, {Smart}, {Souami}, {Spagna},
  {Steele}, {Taris}, {Utrilla}, {van Reeven}, \& {Vecchiato}}]{Lindegren2020}
{Lindegren}, L., {Klioner}, S.~A., {Hern{\'a}ndez}, J., {et~al.} 2020, arXiv
  e-prints, arXiv:2012.03380.
\newblock \doarXiv{2012.03380}

\bibitem[{{Majewski} {et~al.}(2017){Majewski}, {Schiavon}, {Frinchaboy},
  {Allende Prieto}, {Barkhouser}, {Bizyaev}, {Blank}, {Brunner}, {Burton},
  {Carrera}, {Chojnowski}, {Cunha}, {Epstein}, {Fitzgerald}, {Garc{\'\i}a
  P{\'e}rez}, {Hearty}, {Henderson}, {Holtzman}, {Johnson}, {Lam}, {Lawler},
  {Maseman}, {M{\'e}sz{\'a}ros}, {Nelson}, {Nguyen}, {Nidever}, {Pinsonneault},
  {Shetrone}, {Smee}, {Smith}, {Stolberg}, {Skrutskie}, {Walker}, {Wilson},
  {Zasowski}, {Anders}, {Basu}, {Beland}, {Blanton}, {Bovy}, {Brownstein},
  {Carlberg}, {Chaplin}, {Chiappini}, {Eisenstein}, {Elsworth}, {Feuillet},
  {Fleming}, {Galbraith-Frew}, {Garc{\'\i}a}, {Garc{\'\i}a-Hern{\'a}ndez},
  {Gillespie}, {Girardi}, {Gunn}, {Hasselquist}, {Hayden}, {Hekker}, {Ivans},
  {Kinemuchi}, {Klaene}, {Mahadevan}, {Mathur}, {Mosser}, {Muna}, {Munn},
  {Nichol}, {O'Connell}, {Parejko}, {Robin}, {Rocha-Pinto}, {Schultheis},
  {Serenelli}, {Shane}, {Silva Aguirre}, {Sobeck}, {Thompson}, {Troup},
  {Weinberg}, \& {Zamora}}]{APOGEE_2017}
{Majewski}, S.~R., {Schiavon}, R.~P., {Frinchaboy}, P.~M., {et~al.} 2017, \aj,
  154, 94, \dodoi{10.3847/1538-3881/aa784d}

\bibitem[{{McCleery} {et~al.}(2020){McCleery}, {Tremblay}, {Gentile Fusillo},
  {Hollands}, {G{\"a}nsicke}, {Izquierdo}, {Toonen}, {Cunningham}, \&
  {Rebassa-Mansergas}}]{40pcSample2020}
{McCleery}, J., {Tremblay}, P.-E., {Gentile Fusillo}, N.~P., {et~al.} 2020,
  \mnras, 499, 1890, \dodoi{10.1093/mnras/staa2030}

\bibitem[{{Paczynski}(1990)}]{Vmax_Paczynski_1990}
{Paczynski}, B. 1990, \apj, 348, 485, \dodoi{10.1086/168257}

\bibitem[{{Rybizki} {et~al.}(2021{\natexlab{a}}){Rybizki}, {Green}, {Rix},
  {Demleitner}, {Zari}, {Udalski}, {Smart}, \& {Gould}}]{Rybizki2021}
{Rybizki}, J., {Green}, G., {Rix}, H.-W., {et~al.} 2021{\natexlab{a}}, arXiv
  e-prints, arXiv:2101.11641.
\newblock \doarXiv{2101.11641}

\bibitem[{{Rybizki} {et~al.}(2021{\natexlab{b}}){Rybizki}, {Green}, {Rix},
  {Demleitner}, {Zari}, {Udalski}, {Smart}, \& {Gould}}]{Rybizki+21}
---. 2021{\natexlab{b}}, arXiv e-prints, arXiv:2101.11641.
\newblock \doarXiv{2101.11641}

\bibitem[{{Schmidt}(1968)}]{Schmidt1968}
{Schmidt}, M. 1968, \apj, 151, 393, \dodoi{10.1086/149446}

\bibitem[{{Tremblay} {et~al.}(2019){Tremblay}, {Fontaine}, {Fusillo}, {Dunlap},
  {G{\"a}nsicke}, {Hollands}, {Hermes}, {Marsh}, {Cukanovaite}, \&
  {Cunningham}}]{Tremblay2019}
{Tremblay}, P.-E., {Fontaine}, G., {Fusillo}, N. P.~G., {et~al.} 2019, \nat,
  565, 202, \dodoi{10.1038/s41586-018-0791-x}

\bibitem[{{Trumpler} \& {Weaver}(1953)}]{TrumplerWeaver1953}
{Trumpler}, R.~J., \& {Weaver}, H.~F. 1953, {Statistical astronomy}

\bibitem[{{Weidemann}(2000)}]{WD_initial_to_final_mass_ratio_Weidemann_2000}
{Weidemann}, V. 2000, \aap, 363, 647

\bibitem[{{Wojno} {et~al.}(2017){Wojno}, {Kordopatis}, {Piffl}, {Binney},
  {Steinmetz}, {Matijevi{\v{c}}}, {Bland-Hawthorn}, {Sharma}, {McMillan},
  {Watson}, {Reid}, {Kunder}, {Enke}, {Grebel}, {Seabroke}, {Wyse}, {Zwitter},
  {Bienaym{\'e}}, {Freeman}, {Gibson}, {Gilmore}, {Helmi}, {Munari}, {Navarro},
  {Parker}, \& {Siebert}}]{Wojno2017}
{Wojno}, J., {Kordopatis}, G., {Piffl}, T., {et~al.} 2017, \mnras, 468, 3368,
  \dodoi{10.1093/mnras/stx606}

\bibitem[{{Wood}(1992)}]{WD_Galactic_Archeology_Winget_1992}
{Wood}, M.~A. 1992, \apj, 386, 539, \dodoi{10.1086/171038}

\end{thebibliography}
\end{document}